\documentstyle[12pt,preprint,aps,floats,epsfig,amsbsy]{revtex}
\tighten
\newcommand{\bea}{\begin{eqnarray}}
\newcommand{\eea}{\end{eqnarray}}
\newcommand{\beq}{\begin{equation}}
\newcommand{\eeq}{\end{equation}}
\newcommand{\ba}{\begin{array}}
\newcommand{\ea}{\end{array}}
\newcommand{\nn}{\nonumber}
\newcommand{\epsl}{\varepsilon \hspace{-5pt} / }

\newcommand{\rsl}{r \hspace{-5pt} / }
\newcommand{\qsl}{q \hspace{-5pt} / }
\newcommand{\psl}{p \hspace{-5pt} / }

\newcommand{\bra}{\langle}
\newcommand{\ket}{\rangle}
\newcommand{\OEightTree}{\bra s g |O_8|b\ket_{\text{tree}}}

\renewcommand{\a}{\alpha}
\renewcommand{\b}{\beta}
\newcommand{\g}{\gamma}
\renewcommand{\d}{\delta}
\newcommand{\e}{\epsilon}
\newcommand{\p}{\pi}

\newcommand{\ep}{\varepsilon}

\renewcommand{\l}{\lambda}
\newcommand{\m}{\mu}
\newcommand{\n}{\nu}
\newcommand{\G}{\Gamma}
\newcommand{\D}{\Delta}
\renewcommand{\L}{\Lambda}
\renewcommand{\to}{\rightarrow}
\newcommand{\tfrac}[2]{{\textstyle \frac{#1}{#2}}}
\renewcommand{\Re}{\mbox{Re}}
\renewcommand{\Im}{\mbox{Im}}

\newcommand{\xc}{X_{c\hspace{-0.5ex}/}}
\newcommand{\barxc}{\overline{X}_{c\hspace{-0.5ex}/}}
\newcommand{\brc}{\overline{{\cal B}}_{c\hspace{-0.6ex}/}}
\newcommand{\brsg}{\overline{{\cal B}}_{sg}}
\newcommand{\brqqq}{\overline{{\cal B}}_{q'\bar{q}'q}}
\newcommand{\f}[2]{\frac{#1}{#2}}

\begin{document}
\thispagestyle{empty}

\preprint{
\noindent
\hfill
\begin{minipage}[t]{6in}
\begin{flushright}
BUTP--00/22     \\
hep-ph/0009144  \\
\vspace*{1.0cm}
\end{flushright}
\end{minipage}
}

\draft

\title { Calculation of next-to-leading QCD 
corrections to $\boldsymbol{b \to s g}$} 

\author{Christoph Greub and Patrick Liniger\footnote{Work 
partially supported by Schweizerischer
Nationalfonds.}}

\vspace{2.0cm}

\address{
 Institut f\"ur Theoretische Physik, Universit\"at Bern, \\
 CH--3012 Bern, Switzerland} 

\maketitle
\thispagestyle{empty}
\setcounter{page}{0}

\vspace*{1truecm}
\begin{abstract}
In this paper 
a detailed standard model (SM) calculation of 
the $O(\alpha_s)$ virtual corrections to the 
decay width $\G(b \to s g)$ is presented ($g$ denotes a gluon). Also
the complete expressions for the corresponding $O(\alpha_s)$ 
bremsstrahlung corrections to $b \to s g$
are given. 
The combined result 
is free of infrared and collinear singularities, in accordance with
the KLN theorem. Taking into account the existing
next-to-leading logarithmic (NLL) result for the Wilson coefficient
$C_8^{\rm{eff}}$, a complete NLL result for the branching ratio
${\cal B}^{\rm{NLL}}(b \to s g)$ is derived. Numerically, we obtain
${\cal B}^{\rm{NLL}}=(5.0 \pm 1.0) \times 10^{-3}$, which is more than a factor
of two larger than the leading logarithmic result 
${\cal B}^{\rm{LL}}=(2.2 \pm 0.8) \times 10^{-3}$. 
The impact of these corrections on the inclusive charmless hadronic 
branching ratio $\brc$ of $B$-mesons, which can be used to extract 
$|V_{ub}/V_{cb}|$ in the context of the SM, 
is shown to be of similar importance as NLL
corrections to $b$-quark decay modes with three quarks in the final state. 
Finally, the impact of the NLL corrections to $b \to s g$ on $\brc$
is investigated
in scenarios, where the Wilson coefficient $C_8$ is enhanced by new physics.
\end{abstract}

\vfill
%\today

%\pacs{Preliminary Version}

\setlength{\parskip}{1.2ex}

\section{Introduction}
\label{intro}
The theoretical predictions for inclusive decay rates of $B$-mesons
rest on solid grounds due to the fact that these rates
can be systematically expanded in powers of $\L_{\rm QCD}/m_b$
\cite{Bigi1,Bigi2}, where the leading term corresponds to the decay width
of the underlying $b$-quark decay. 
As the power corrections only start at $O(\L^2_{\rm{QCD}}/m_b^2)$,
they affect these rate by at most a few percent. Theoretically, spectator
effects of order $16\pi^2(\L_{\rm{QCD}}/m_b)^3$ 
\cite{Bigi3,Sachrajda} could be larger \cite{Sachrajda}, 
but for the decay rates of $B^\pm$ and $B^0$
they are experimentally known to be at the percent level
as well \cite{Kroll}. 
Thus 
the accuracy of the theoretical predictions is mainly controlled 
by our knowledge
of the perturbative corrections to the free quark decays.

The inclusive charmless hadronic decays, $B \to \xc$, 
where $\xc$ denotes any hadronic 
charmless final state, are an interesting subclass of the decays mentioned 
above; as pointed out in ref. \cite{Lenz2}, a measurement of 
the corresponding branching ratio would allow the extraction of
the presently poorly known 
ratio $|V_{ub}/V_{cb}|$, where $V_{ub}$ and $V_{cb}$ are elements
of the Cabibbo-Kobayashi-Maskawa (CKM) matrix.
At the quark level, there are decay modes with three-body final states,
viz. $b \to q' \overline{q}' q$, ($q'=u,d,s$; $q=d,s$) and the modes
$b \to q g$, with two-body final state topology, which contribute 
to the charmless decay width at leading logarithmic (LL) accuracy.
Calculations of
next-to-leading logarithmic (NLL) corrections to the three-body decay modes
were started already some time ago in ref. \cite{Altarelli}, where
radiative corrections to the current-current diagrams of the operators
$O_1$ and $O_2$ were calculated, together with NLL corrections to the
Wilson coefficients. Later, Lenz. et al. included in a first step
\cite{Lenz1} the contributions of the penguin diagrams associated with
the operators $O_1$ and $O_2$, and in a second step 
\cite{Lenz2} the same authors also
included one-loop penguin diagrams of the penguin operators $O_3,...,O_6$;
also the effects of the chromomagnetic operator $O_8$ were taken
into account to the relevant precision needed for a NLL calculation. 
Up to contributions from current-current type corrections to the penguin
operators, the NLL calculation for the three quark final states is now 
complete. 

In the numerical evaluations of the charmless hadronic branching ratio, 
the two body decay modes $b \to q g$
were added in refs. \cite{Lenz1,Lenz2} at the LL precision, 
as the full NLL predictions were missing. To fill this gap, we 
recently wrote a short letter where {\it NLL results} for the branching ratio
${\cal B}(b \to s g)$ were presented \cite{Liniger}. 
In the present work, we describe in detail the non-trivial two-loop 
{\it NLL calculation}, which led to the results in \cite{Liniger}.
As the NLL corrections enhance ${\cal B}(b \to s g)$ by more than a factor
of 2, we also analyze in the present paper their impact on the 
charmless hadronic branching ratio.

The decay $b \to s g$ gained a lot of attention in the last years. For a long
time the theoretical predictions for both, the inclusive
semileptonic branching ratio ${\cal B}_{\rm{sl}}$
and the charm multiplicity $n_c$ in $B$-meson
decays were considerably higher than the experimental values
\cite{Bigi_Falk}. 
An attractive
hypothesis, which would move the theoretical predictions
for both observables into the direction favored by the experiments,
assumed the rare decay mode $b \to s g$ to be enhanced by
new physics.

After the inclusion of the complete NLL corrections to the decay modes
$b \to c \overline{u} q$ and $b \to c \overline{c} q$ ($q=d,s$) \cite{Bagan},
the theoretical prediction for the semileptonic branching ratio
and the charm multiplicity \cite{Sachrajda}
are
\beq
{\cal B}_{\rm{sl}}^{\rm{th}}=(11.7 \pm 1.4 \pm 1.0) \% \ , \quad
n_c^{\rm{th}}=1.20 \pm 0.06 \ , 
\eeq
where the second error in ${\cal B}_{\rm{sl}}^{\rm{th}}$ takes into account
the spectator effects estimated in ref. \cite{Sachrajda}.
The experimental results from measurements at the $\Upsilon(\rm{4S})$
resonance and those from the $Z^0$ resonance at LEP and SLD
were recently summarized \cite{Japantalk} to be
\bea
&&
{\cal B}_{\rm{sl}}^{\Upsilon\rm{(4S)}}=(10.45 \pm 0.21) \% \ , \quad
n_c^{\Upsilon\rm{(4S)}}=1.14 \pm 0.06 \ , \nn \\ 
&&
{\cal B}_{\rm{sl}}^{Z^0}=(10.79 \pm 0.25) \% \ , \quad
n_c^{Z^0}=1.17 \pm 0.04 \ . 
\eea
We would like to stress that in the theoretical results the renormalization
scale was taken in the interval $[m_b/4,2m_b]$. If one only considers
$\mu \in [m_b/2,2m_b]$, the theoretical predictions
would only have marginal
overlap with
experimental data. This implies that 
there is still room for enhanced $b \to s g$. We therefore also
illustrate in this paper the influence of the NLL corrections
to $b \to s g$ on the charmless hadronic branching ratio in scenarios where
the Wilson coefficient $C_8$ is enhanced by new physics.

We also would like to mention that the component $b \to s g$
of the charmless hadronic decays is expected to manifest itself in 
kaons with high momenta (of order $m_b/2$), due to its two body nature
\cite{Rathsman}. Some indications for enhanced $b \to s g$
in this context were reported by
the SLD collaboration \cite{SLD}. For a review of other hints for
enhanced $b \to s g$, the reader is referred to \cite{Kagan_hawaii}.   

Within the SM, the LL prediction for the branching for $b \to s g$
is known to be ${\cal B}(b \to s g) \approx 0.2\%$ \cite{Ciuchini}.
The process $b \to s g g$, which gives a NLL contribution to the
inclusive charmless decay width has already been studied in the literature
\cite{Hou,Simma}. In \cite{Simma} a complete calculation was performed
in regions of the phase space which are free of collinear an infrared 
singularities. Putting suitable cuts, the branching ratio for $b \to s g g$
was found to be of the order $10^{-3}$ in these phase space regions.
A complete calculation requires the calculation of a regularized version
for the decay width $\G(b \to s g g)$ in which infrared and collinear
singularities become manifest. Only after adding the virtually corrected
decay width $\G(b \to s g)$ a meaningful physical result is obtained.
In addition, as we will see later, also the tree level contribution of the
operator $O_8$ to the decays $b \to s f \overline{f}$, with
$f=u,d,s$, has to be included. 

The remainder of this paper is organized as follows: In section 
\ref{effective_Ham}, we review the theoretical framework and discuss
the steps needed for a NLL calculation for ${\cal B}(b \to s g)$. 
Section \ref{virtcorr} is devoted to the calculation 
of the virtual corrections to the matrix elements $\bra s g|O_{1,2}|b \ket$,
including renormalization, while section \ref{virto8} deals with virtual
corrections to  $\bra s g|O_{8}|b \ket$. 
In section \ref{Gammavirt} the virtual corrections to the decay width
$\G(b \to s g)$ are calculated.
In section \ref{Quarkstrahlung} the decay width $\G(b \to s f \overline{f})$
is given. Sections \ref{bremsmat} and \ref{bremswidth} deal with the gluon
bremsstrahlung matrix elements $\bra s g g|O_{1,2,8}|b \ket $ and the
corresponding decay width, respectively. The analytic results for the NLL
branching ratio ${\cal B}(b \to s g)$ can be found in section \ref{brnll}, 
while the numerical evaluations are presented in section \ref{numres}.
Section \ref{charmless} deals with the impact of the NLL corrections
to ${\cal B}(b \to s g)$ on the charmless hadronic branching ratio in the
standard model, while in section \ref{c8enh} similar questions are 
addressed in scenarios where the Wilson coefficient $C_8$ is enhanced by new
physics. We conclude with a short summary in section \ref{summary}.
An explicit parametrization of the NLL Wilson coefficient 
$C_8^{\rm{eff}}(m_b)$ is given in  appendix \ref{appendix:a}.

\section{The effective Hamiltonian}
\label{effective_Ham}

\noindent 
We use the framework of an effective low--energy theory with five
quarks, obtained by integrating out the heavy degrees of freedom,
which in the SM are the $t$--quark and the $W$--boson.
We  take
into account operators up to dimension six and we put $m_s=0$.
In  this approximation the   
effective Hamiltonian relevant for radiative decays and $b \to s g(g)$ 
reads  
\beq
\label{heff}
{\cal H}_{\rm eff} = - \frac{4 G_F}{\sqrt{2}} \,V_{ts}^* V_{tb} 
   \sum_{i=1}^8 C_i(\mu) O_i(\mu),  
\eeq
where $G_F$ is the Fermi coupling constant and $C_i(\mu)$ are
the Wilson coefficients evaluated at the scale $\mu$; 
$V_{tb}$ and $V_{ts}$ are matrix elements of the 
Cabibbo-Kobayashi-Maskawa
(CKM) matrix.
The operators $O_i$ read \cite{Misiak97}
\beq
\begin{array}{llll}
O_1 \,= &\!
 (\bar{s}_L \gamma_\mu T^A c_L)\, 
 (\bar{c}_L \gamma^\mu T^A b_L)\,, 
               &  \quad 
O_2 \,= &\!
 (\bar{s}_L \gamma_\mu c_L)\, 
 (\bar{c}_L \gamma^\mu b_L)\,,   \\[1.002ex]
O_3 \,= &\!
 (\bar{s}_L \gamma_\mu b_L) 
 \sum_q
 (\bar{q} \gamma^\mu q)\,, 
               &  \quad 
O_4 \,= &\!
 (\bar{s}_L \gamma_\mu T^A b_L) 
 \sum_q
 (\bar{q} \gamma^\mu T^A q)\,,  \\[1.002ex]
O_5 \,= &\!
 (\bar{s}_L \gamma_\mu \gamma_\nu \gamma_\rho b_L) 
 \sum_q
 (\bar{q} \gamma^\mu \gamma^\nu \gamma^\rho q)\,, 
               &  \quad 
O_6 \,= &\!
 (\bar{s}_L \gamma_\mu \gamma_\nu \gamma_\rho T^A b_L) 
 \sum_q
 (\bar{q} \gamma^\mu \gamma^\nu \gamma^\rho T^A q)\,,  \\[1.002ex]
O_7 \,= &\!
  \frac{e}{16\pi^2} \,{\overline m}_b(\mu) \,
 (\bar{s}_L \sigma^{\mu\nu} b_R) \, F_{\mu\nu}\,, 
               &  \quad 
O_8 \,= &\!
  \frac{g_s}{16\pi^2} \,{\overline m}_b(\mu) \,
 (\bar{s}_L \sigma^{\mu\nu} T^A b_R)
     \, G^A_{\mu\nu} \ .
\end{array} 
\label{opbasis}
\eeq
In the dipole operators $O_7$ ($O_8$), $e$ and $F_{\m \n}$
($g_s$ and $G_{\m \n}^A$) denote the electromagnetic (strong)
coupling constant and field strength tensor, respectively.
$T^A$ ($A=1,...,8$) are $SU(3)$ color generators;
$L=(1-\gamma_5)/2$ and $R=(1+\gamma_5)/2$ stand for left- and right-handed
projectors.
In eq.~(\ref{opbasis}), 
${\overline m}_b(\mu)$ is the running $b$--quark mass 
in the ${\overline{\mbox{MS}}}$ scheme at the renormalization scale $\mu$. 
Henceforth, ${\overline m}_q(\mu)$ and $m_q$ denote ${\overline{\mbox{MS}}}$ 
running and pole masses, respectively. To first order in 
$\a_s$, these masses are related through:
\beq
 {\overline m}_q(\mu) = {m_q} 
 \left(1 + \frac{\a_s(\mu)}{\pi} \, \ln \frac{m_q^2}{\mu^2} 
- \frac{4}{3} \frac{\a_s(\mu)}{\pi}\right) \, .
\label{polerunning}
\eeq

It is well-known that QCD corrections to the decay rate 
for $b \to s \gamma$ bring in logarithms of the mass ratios $m_b/m_W$
and $m_b/m_t$.
The same is true for the process $b \to s g$: QCD corrections to this process
induce terms of the form $\a_s \a_s^n \ln^m(m_b/M)$, where $M=m_t$ or $m_W$
and $m \le n$ (with $m,n=0,1,2,\ldots$).

One can systematically resum these large terms by renormalization
group techniques. Usually, one matches the full standard model theory
with the effective theory at a scale of order $m_W$. At this scale,
the large logarithms generated by matrix elements in the 
effective theory are the  same ones
as in the full theory. Consequently, the
Wilson coefficients only pick up formally small QCD corrections.
Using the renormalization group equation, the Wilson coefficients
are then calculated at the scale $\mu=\mu_b \approx m_b$, at which
the large logarithms are contained in 
the Wilson coefficients, while the matrix elements of the 
operators are free of them.

So far the decay rate for $b \to s g$ has been 
systematically
calculated only to leading logarithmic (LL) accuracy, i.e., for $m=n$.

A consistent calculation for 
$b \to s g$ at LL
precision requires the following steps:
%\\{\it 1)}
\begin{itemize}
\item[{\it 1)}] 
the extraction of the Wilson coefficients from 
a matching calculation of the full standard model theory 
with the effective theory at the scale $\mu=\mu_W$ 
to order $\alpha_s^0$;
$\mu_W$ denotes a scale of order $m_W$ or $m_t$;
%\\{\it 2)}
\item[{\it 2)}]  
a renormalization group treatment of the Wilson coefficients,
using the anomalous-dimension matrix to order $\alpha_s^1$;
%\\{\it 3)}
\item[{\it 3)}]   
the calculation of the decay matrix elements $\bra s g |C_i O_i|b \ket$
at the scale 
$\mu = \mu_b$  to order $g_s$; $\mu_b$ denotes a scale of order $m_b$.
We note that the matrix elements associated with the four Fermi operators
($i=1-6$) can be absorbed into the effective Wilson coefficient 
$C_8^{\rm{eff}}$, when working at LL precision. In the naive dimensional
regularization scheme (NDR), 
which we use in this paper, 
one obtains \cite{Misiak97}
\beq
\label{C8eff}
C_8^{\rm{eff}} =  C_8 + C_3 - \tfrac{1}{6} \, C_4 + 20 \, C_5 - 
\tfrac{10}{3} \, C_6 \quad .
\eeq
\end{itemize}

{}From the analogous decay $b \to s \gamma$ it is well-known
that next-to-leading logarithmic (NLL) corrections 
drastically reduce the large renormalization scale dependence
of the LL branching ratio. This implies, in particular,
that the NLL corrections are relatively large, at least
for certain scales (within the usually considered range $
m_b/2 \le \mu_b \le 2 m_b$). 
Motivated by the situation in this analogous
process, we present in this paper a systematic 
calculation of the NLL corrections to $b \to s g$.
 
The principal organization of such a calculation is straightforward:
Each of the three steps listed above has to be improved by going 
to the next order in $\a_s$: {\it 1)} The matching has to be calculated 
including $\a_s$ corrections; {\it 2)} 
the renormalization group treatment of the Wilson 
coefficients has to be performed using the anomalous dimension matrix
to order $\a_s^2$; {\it 3)}
finally, the order $\a_s$ corrections to the decay matrix
elements have to be worked out. We note that this step involves
both, the calculation of {\it virtual-} and {\it bremsstrahlung} corrections
to $b \to s g$. 

The first two steps are already available in the literature.
The order $\a_s$ matching of the dipole operators $O_7$ and
$O_8$ was calculated in refs. \cite{Adel},
while the matching conditions and the anomalous dimension matrix 
for the four Fermi operators have been calculated by several groups 
\cite{anomal_old}. These calculations were done in the ``old 
operator basis'', introduced by Grinstein et al. \cite{Grinstein90}.
The most difficult part, the order $\a_s^2$ mixing
of the four-Fermi operators into the dipole operators 
requires the calculation of three loop diagrams
\cite{Misiak97}.
In order to perform a consistent NDR  calculation (i.e. with anticommuting 
$\gamma_5$), the old operator basis was replaced 
by the new one displayed in eq. (\ref{opbasis}). The full $8\times 8$
anomalous dimension matrix, the corresponding matching conditions and the
definition of the evanescent operators is given in ref. \cite{Misiak97}
and is repeated in appendix \ref{appendix:a} of the present paper. 

Step {\it 3)}, the calculation of the virtual $O(\a_s)$ 
corrections to the matrix
elements $M_i=\bra s g |O_i|b \ket$, as well as the evaluation of the
gluon bremsstrahlung process $b \to s g g $, is performed the first time
in the present paper.  
As illustrated in table \ref{table1}, the LL Wilson coefficients
$C_3^0(\mu_b)$,...,$C_6^0(\mu_b)$ 
are much smaller than $C_1^0(\mu_b)$ and $C_2^0(\mu_b)$. 
We therefore only calculate
$M_1$, $M_2$ and $M_8$ together with the corresponding bremsstrahlung
corrections.
As $M_1$ and $M_2$ vanish at
one-loop (i.e. without QCD corrections), only the leading order pieces,
$C_1^0(\mu_b)$ and $C_2^0(\mu_b)$, appearing in the decomposition 
\beq
\label{wilsondecomp}
C_i(\m_b) = C_i^0(\m_b) + \frac{\a_s(\m_b)}{4\p} \, C_i^1(\m_b)
\eeq
of the NLL Wilson coefficients $C_1(\mu_b)$ and
$C_2(\mu_b)$ are needed.
On the other hand, the operator $O_8$ contributes to $M_8$ already
at tree-level. Consequently the full NLL Wilson coefficient
$C_8^{\rm{eff}}(\mu_b)$ is needed.
The numerical value of the NLL 
piece $C_8^{1,\rm{eff}}$ (defined as in eq. (\ref{C8eff}) and (\ref{wilsondecomp})) 
is also given in table \ref{table1}, while the 
analytic form is relegated to appendix \ref{appendix:a}.

\begin{table}[htb]
\label{coeff}
\begin{center}
\begin{tabular}{| c | c | c | c | c | }
%\hline
  & $\mu=m_W$ & $\mu=9.6$ GeV
                              & $\mu=4.8$ GeV
                              & $\mu=2.4$ GeV\\
\hline \hline
$\a_s $ & $0.121$ & $0.182$ & $0.218$ & $0.271$ \\
\hline
$C_1^0$ & $0.0$ & $-0.335$ & $-0.497$ & $-0.711$ \\
$C_2^0$ & $1.0$ & $1.012$ & $1.025$ & $1.048$ \\
$C_3^0$ & $0.0$ & $-0.002$ & $-0.005$ & $-0.010$ \\
$C_4^0$ & $0.0$ & $-0.042$ & $-0.067$ & $-0.103$ \\
$C_5^0$ & $0.0$ & $0.0002$ & $0.0005$ & $0.001$ \\
$C_6^0$ & $0.0$ & $0.0005$ & $0.001$ & $0.002$ \\
$C_7^0$ & $-0.192$ & $-0.285$ & $-0.324$ & $-0.371$ \\
$C_8^0$ & $-0.096$ & $-0.136$ & $-0.150$ & $-0.166$ \\
$C_7^{0,\rm{eff}}$ & $-0.196$ & $-0.280$ & $-0.314$ & $-0.356$ \\
$C_8^{0,\rm{eff}}$ & $-0.097$ & $-0.135$ & $-0.149$ & $-0.165$ \\
$C_8^{1,\rm{eff}}$ & $-2.166$ & $-1.318$ & $-1.098$ & $-0.950$ \\
$C_8^{\rm{eff}}$ & $-0.118$ & $-0.154$ & $-0.168$ & $-0.186$ \\
%\hline
\end{tabular}
\end{center}
\caption{Wilson coefficients $C_i^0(\mu )$ ($i=1,...,8$),
$C_8^{1,\rm{eff}}$ and  $C_8^{\rm{eff}}$
(see eq. (\ref{wilsondecomp}) in the text)
at the matching scale $\mu=m_W=80.33$ GeV
and at three other scales, $\mu = 9.6$ GeV,
$\mu =4.8$ GeV and $\mu = 2.4$ GeV. For $\a_s(\mu)$
(in the $\overline{\mbox{MS}}$ scheme) we used the
two-loop expression with 5 flavors and $\a_s(m_Z)=0.119$.
The entries correspond to the pole top quark mass
$m_t= 175 $ GeV.}
\label{table1}
\end{table}

%%%%%%%%%%%%%%%%%%%%%% SECTION 3 %%%%%%%%%%%%%%%%%%%%%%%%%%%%%%%%
\section{Virtual corrections to $\boldsymbol{O_1}$ and $\boldsymbol{O_2}$}
\label{virtcorr}

In this section
we present the calculation of the matrix elements of the operator 
$O_1$ and $O_2$
for $b \to s g$ up to order $\a_s$ in the NDR scheme. 
The one-loop
$(\a_s^0)$  matrix elements vanish and we must consider
several two-loop contributions.
Since they involve ultraviolet singularities also
counterterm contributions are needed. These are
easy to obtain, because  the operator renormalization constants
$Z_{ij}$ are known with enough
accuracy from the order $\alpha_s$
anomalous dimension matrix \cite{Misiak97}.
%Explicitly,
%we need the contributions of the operators
%$C_1 \delta Z_{1j} O_j$ and
%$C_2 \delta Z_{2j} O_j$ 
%to the matrix elements for
%$b \to s g$,
%where $\delta Z_{1j}$ and $\delta Z_{2j}$ 
%denote the order $\alpha_s$
%contribution of the operator renormalization constants.
%In the NDR scheme, the non-vanishing
%counterterms come from the one-loop matrix
%elements of $C_1 \delta Z_{14} O_4$ and
%$C_2 \delta Z_{24} O_4$ as well as from the tree level
%matrix elements of the operators 
%$C_1 \delta Z_{18} O_8$ and
%$C_2 \delta Z_{28} O_8$.
%We also note that there are no contributions to $b \to s g$
%from counterterms proportional to evanescent 
%operators multiplying the Wilson coefficients $C_1$ and $C_2$.

\subsection{Regularized two-loop matrix elements
 of $\boldsymbol{O_1}$ and $\boldsymbol{O_2}$}
For the following discussion it is useful to define the operators
$\hat{O}_1$ and $\hat{O}_2$:
\beq
\hat{O}_1 = 2 O_1 + \tfrac{1}{3} O_2 \quad ; \quad
\hat{O}_2 = O_2 \quad .
\label{o12def}
\eeq
$\hat{O}_1$ and $\hat{O}_2$ are nothing but the current-current operators in
the old basis \cite{Grinstein90}:
\beq
\begin{array}{llll}
\hat{O}_1 \,= &\!
 (\bar{s}_{L \a} \gamma_\mu  c_{L \b})\, 
 (\bar{c}_{L \b} \gamma^\mu  b_{L \a})\,, 
               &  \quad 
\hat{O}_2 \,= &\!
 (\bar{s}_{L \b} \gamma_\mu c_{L \b})\, 
 (\bar{c}_{L \a} \gamma^\mu b_{L \a})\,.   \\[1.002ex]
\end{array} 
\label{o12}
\eeq
 
\begin{figure}[t]
\begin{center}
\leavevmode
\includegraphics[scale=1.2]{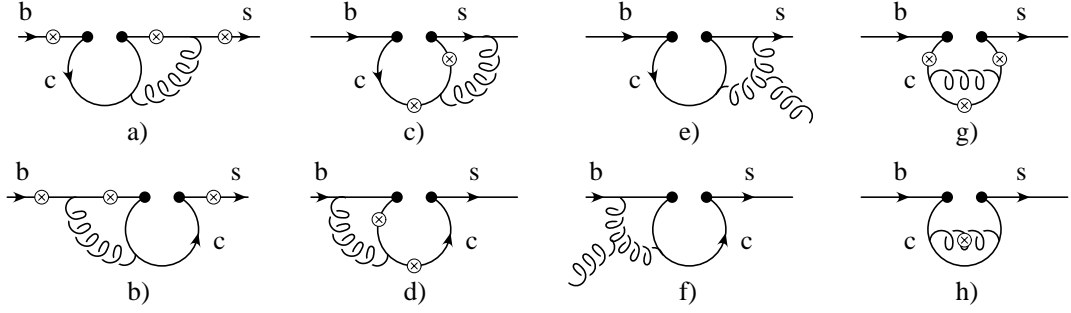}
\end{center}
\caption[f1]{Graphs associated with the operators $\hat{O}_1$ and $\hat{O}_2$.
The wavy lines represent gluons; the real gluons are understood to be
attached to the circle-crosses.}
\label{fig:1new}
\end{figure}

We now present the calculation of the matrix elements
$\hat{M}_i = \bra s g |\hat{O}_i | b \ket$:
The dimensionally regularized matrix element 
$\hat{M}_2$
is obtained by calculating the two-loop diagrams a) -- h)
shown in fig. \ref{fig:1new}.

\begin{figure}[t]
\begin{center}
\leavevmode
\includegraphics[scale=1.2]{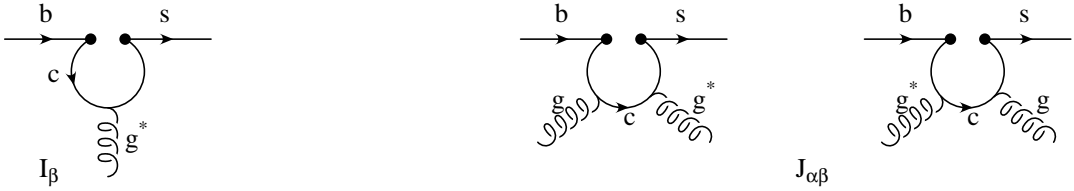}
\end{center}
\caption[f1]{Building block $I_\b$ 
(with an off-shell gluon) 
for the diagrams a), b), e) and f) in
  fig. \ref{fig:1new}
and building block $J_{\a\b}$
for the diagrams c) and d) in fig. \ref{fig:1new}.
$g^*$ and $g$ denote an off-shell and an on-shell gluon,
respectively.}
\label{fig:2new}
\end{figure}

We start with the calculation of the diagrams a) -- f) in fig. \ref{fig:1new},
in which the virtual gluon connects the charm quark in the loop with an
external fermion leg\footnote{
The diagrams g) and h)  are much easier to
calculate than those in a) -- f), because
$m_c$ is the only scale in the corresponding integrals.}.
The main steps of the calculation are the following: We first
calculate the Fermion loops in the individual diagrams,
i.e.,  the 'building blocks' $I_\b$ and $J_{\a \b}$ shown in
fig. \ref{fig:2new}; $J_{\alpha\beta}$ denotes the 
sum of the the two diagrams
on the right.

We work in $d=4-2 \e$
dimensions; the results of the building blocks are presented as integrals over
Feynman parameters after
integrating over the (shifted) loop-momentum.
Then we insert these
building blocks into the full two-loop diagrams.
Using one more Feynman parametrization,
we calculate the integral over the second loop-momentum.
As the remaining Feynman parameter integrals contain rather
complicated denominators, we do not evaluate them directly.
At this level we also do not expand in the regulator $\e$.
The heart of our procedure
which will be explained more explicitly below,
is to
represent these denominators as  complex Mellin-Barnes integrals
 \cite{Abra}. After inserting this
representation and interchanging the order of integration, the
Feynman parameter integrals are reduced to well-known Euler
Beta-functions. Finally, the residue theorem
allows  to write the result of the remaining
complex integral as the sum over the residues taken at
the pole positions of
Beta- and
Gamma-functions; this naturally leads
to an expansion in the ratio $z=(m_c/m_b)^2$, which
numerically is about $z=0.1$.

We express 
the diagram on the left in fig. \ref{fig:2new} (denoted by $I_\b$)
in a way convenient for
inserting into the two-loop
diagrams. As we will use $\overline{\mbox{MS}}$ subtraction later on,
we introduce the renormalization scale in the form
$\mu^2 \exp(\gamma_E) /(4 \pi)$, where
$\g_E \approx 0.577$ is the Euler constant.
Then,
$\overline{\mbox{MS}}$ corresponds to subtracting
the poles in $\e$.
In the NDR scheme, $I_\beta$ is given 
by\footnote{The fermion/gluon and the fermion/photon couplings
are defined according to the covariant derivative
$D = \partial + i g_s T^B  A^B + i e Q A$ where $T^B = \frac{\lambda^B}{2}$
are the SU(3) generators.}
\bea
\label{build1}
I_\beta^A &=& - \frac{g_s}{4 \pi^2} \, \Gamma(\e) \,
\mu^{2 \e} \,
\exp(\gamma_E \e)
\, (1-\e)
\exp(i \pi \e) \,T^A  \left(r_\b \rsl - r^2 \g_\b \right) \, L
\times \nonumber \\
&& \int_0^1 [x(1-x)]^{1-\e} \,
\left[ r^2 - \frac{m_c^2}{x(1-x)} + i \delta \right]^{-\e}
\quad ,
\eea
where $r$ is the four-momentum of the (off-shell) gluon,
$m_c$ is the mass of the charm quark propagating in the loop
and the term $i \delta$ is the "$\e$-prescription".
The free index $\b$ will be contracted with
the gluon propagator when inserting the building block into
the two-loop diagrams a), b), e) and f) in fig. \ref{fig:1new}.
Note that $I_\beta$ is gauge invariant
in the sense that $r^\beta I_\beta = 0$.

Next we give the sum of the two diagrams on the right in fig. \ref{fig:2new},
using the decomposition in \cite{Simma}. The on-shell gluon has momentum
$q$, color $A$ and polarization $\a$ (therefore we drop the terms $q^2$ and
$q_\a$), while the off-shell gluon has momentum $r$, color $B$ and
polarization $\b$.  This building block, denoted by $J_{\a\b}^{AB}$, can be
decomposed with respect to the color structure as \beq
\label{build2}
J_{\a \b}^{AB} = T^+_{\a \b}(q,r) \left\{ T^A , T^B \right\} + T^-_{\a
  \b}(q,r) \left[ T^A , T^B \right] \quad .  
\eeq 
The quantities $T^+_{\a \b}(q,r)$ and $T^-_{\a \b}(q,r)$ read
\bea
\label{tp}
T^+_{\a \b}(q,r)  &=& \frac{g^2_s}{32 \pi^2} \,
\bigg[  E(\a,\b,r)  \, \Delta i_5
       + E(\a,\b,q) \, \Delta i_6
       - E(\b,r,q) \, \frac{r_\a}{(qr)} \Delta i_{23}
       \nonumber  \\ &&  \hspace{1.0cm}
       - E(\a,r,q) \, \frac{r_\b}{(qr)} \Delta i_{25}
       - E(\a,r,q) \, \frac{q_\b}{(qr)} \Delta i_{26}
       \bigg] \, L  \quad ,\\
\label{tm}
T^-_{\a \b}(q,r)  &=& \frac{g^2_s}{32 \pi^2} \,
\bigg[  
\rsl \, g_{\a \b} \, \Delta i_2
+ \qsl \, g_{\a \b} \, \Delta i_3
+  \g_\b \, r_\a \, \Delta i_8
+  \g_\a \, r_\b \, \Delta i_{11}
+  \g_\a \, q_\b \, \Delta i_{12}
       \nonumber  \\ &&  \hspace{1.0cm}
+ \rsl \, \frac{r_\a r_\b}{(qr)} \, \Delta i_{15}
+ \rsl \, \frac{r_\a q_\b}{(qr)} \, \Delta i_{17}
+ \qsl \, \frac{r_\a r_\b}{(qr)} \, \Delta i_{19}
+ \qsl \, \frac{r_\a q_\b}{(qr)} \, \Delta i_{21}
       \bigg] \, L  \quad .
\eea
The matrix $E$  in eq.
(\ref{tp}) is defined as
\beq
\label{epsilongeneralization}
E(\a,\b,r) = \g_\a \g_\b \rsl - \g_\a r_\b + \g_\b (r_\a)
- \rsl g_{\a\b} .
\eeq
In a four-dimensional
context these $E$ quantities can be reduced to expressions
involving the
Levi-Civit\`a tensor, i.e., $E(\a,\b,\g) = -i \,
\ep_{\a \b \g \mu} \, \g^\mu \g_5 $ (in the Bjorken-Drell
convention).
The dimensionally regularized expressions for the
$\Delta i$ functions read
\bea
\label{deltai5}
\Delta i_{5\phantom{2}}  & = & - 4 \, B^+ \int_S \, dx \, dy
\, C^{-1-\e}  \big[
4 (qr) x^2 y \e - 4 (qr) x y \e 
 \nn \\  & & \qquad \qquad \qquad 
 - 2 r^2 x^3  \e + 3 r^2 x^2 \e
- r^2 x \e + 3 x C - C \big]  \\
\label{deltai6}
\Delta i_{6\phantom{2}} & = & \phantom{-} 4 \, B^+ \int_S \, dx \, dy
\,C^{-1-\e} \big[
4 (qr) x y^2 \e - 4 (qr) x y \e
\nn \\  & & \qquad \qquad \qquad 
- 2 r^2 x^2 y \e
+ 2 r^2 x^2 \e
+ r^2 x y \e 
- 2 r^2 x \e + 3 y C - C \big]   \\
\label{deltai23}
\Delta i_{23} & = & -\Delta i_{26} = 8 \, B^+ (qr) \e \int_S \, dx \, dy \, C^{-1-\e} x y \\
\label{deltai25}
\Delta i_{25} & = & -8 \, B^+ (qr) \e \int_S \, dx \, dy \, C^{-1-\e}
x(1-x)
\eea
\bea
\label{deltai2}
\Delta i_{2\phantom{1}} &=& \phantom{-} 4 \, B^-  \int_S \, dx \, dy
\,  C^{-1-\e} (1-x) \, \left[
4 (qr) x y \e - 2 r^2 x^2 \e + r^2 x \e
+C \right] \\
\label{deltai3}
\Delta i_{3\phantom{1}} &=& \phantom{-} 4 \, B^-  \int_S \, dx \, dy
\, C^{-1-\e} \big[
4 (qr) x y^2 \e - 4 (qr) x y \e 
\nn \\ && \qquad \qquad \qquad
- 2 r^2 x^2 y \e 
+ 2 r^2 x^2 \e
+ r^2 x y \e - 2 r^2 x \e + y C - C \big]  \\
\label{deltai8}
\Delta i_{8\phantom{1}} &=& -4 \, B^-  \int_S \, dx \, dy
\,  C^{-1-\e} \big[
4 (qr) x^2 y \e + 2 (qr) x y \e 
 \nn \\ && \qquad \qquad \qquad
- 2 r^2 x^3 \e  
+  r^2 x^2 \e
+ r^2 x  \e + x C + C \big] \\
\label{deltai11}
\Delta i_{11} &=& \phantom{-} 4 \, B^-  \int_S \, dx \, dy
\, C^{-1-\e} (1-x) \, \left[
4 (qr) x y \e - 2 (qr) x  \e - 2 r^2 x^2 \e +  r^2 x \e
+ C \right]  \\
\label{deltai12}
\Delta i_{12} &=& \phantom{-} 4 \, B^-  \int_S \, dx \, dy
\,  C^{-1-\e} \big[
4 (qr) x y^2 \e + 2 (qr) x y \e 
 \nn \\ && \qquad \qquad \qquad
- 2 r^2 x^2 y \e 
- 2 r^2 x^2 \e
+ r^2 x y \e + 2 r^2 x \e + y C + C \big] \\
\label{deltai15}
\Delta i_{15} &=& \phantom{l} 16 \, B^- (qr) \e \int_S \, dx \, dy
\,  C^{-1-\e}  x^2 (1-x) \\
\label{deltai17}
\Delta i_{17} &=& -8 \, B^-  (qr)\e  \int_S \, dx \, dy
\,  C^{-1-\e} x y (1-2x) \\
\label{deltai19}
\Delta i_{19} &=& \phantom{-} 8 \, B^-  (qr) \e \int_S \, dx \, dy
\, C^{-1-\e}  x (1-x-y+2xy) \\
\label{deltai21}
\Delta i_{21} &=& \phantom{-} 8 \, B^-  (qr) \e \int_S \, dx \, dy
\,  C^{-1-\e} x y (1-2y)
\eea
where $C$, $C^{-1-\e}$ and $B^\pm$ are given by
\bea
\label{cs}
C &=& m_c^2 - 2 x y (qr) -  x (1-x) r^2 -i \delta \nonumber \\
C^{-1-\e} &=& - \exp(i \pi \e) \,  [x(1-x)]^{-1-\e} \,
\left[ r^2 + \frac{2 y (qr)}{1-x} - \frac{m_c^2}{x(1-x)} + i \delta
\right]^{-1-\e}
\eea
\begin{eqnarray}
  \label{eq:bs}
  B^+ =  (1+\e)\,\Gamma(\e) \exp(\gamma_E \e)
               \mu^{2\e} \, ,
  & \qquad \qquad & B^- =  (\e-1)
\Gamma(\e) \exp(\gamma_E \e) \mu^{2\e} \quad .
\end{eqnarray}
The range of integration in $(x,y)$ is restricted to the
simplex $S$, i.e.,  $0 \le y \le (1-x)$ and $0 \le x \le 1$.

We are now ready to evaluate the two-loop diagrams.
%As both  $I_\b$ and $J_{\a\b}$
%are transverse with respect to
%the gluon, the gauge of the gluon propagator
%is irrelevant.
Due to the absence of extra singularities in the limit of
vanishing strange quark mass, we set $m_s=0$ from
the very beginning.

\begin{figure}[h]
  \begin{center}
    \leavevmode
    \includegraphics[scale=0.3]{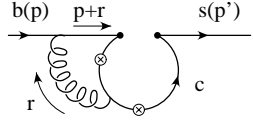}
  \end{center}
  \caption[f1]{Feynman diagram for the Mellin-Barnes example. The momentum
and the polarization vector of the emitted gluon are denoted by $q$
and $\varepsilon$, respectively.}
  \label{fig:1Momenta}
\end{figure}

In ref. \cite{GHW} the detailed calculation of one of the diagrams
in fig. \ref{fig:1new}a)
was presented for $b \to s \g$.
As all the other diagrams, which involve the building block $I_\b$,
i.e., a), b), e) and f) in fig. \ref{fig:1new},
can be computed in a very similar way, we prefer to concentrate
on the diagrams involving the building block $J_{\a \b}$.
As an example in this class, we concentrate on the diagrams d) in fig.
\ref{fig:1new}, which we redisplay in fig. \ref{fig:1Momenta} in order to
set up the notation for the momenta.

The sum $\hat{M}_2(d)$ of the two diagrams can be decomposed 
into a color symmetric part $\hat{M}_2^+(d)$ and a color antisymmetric part
$\hat{M}_2^-(d)$ according to
\beq
\hat{M}_2(d) = \hat{M}_2^+(d) + \hat{M}_2^-(d) \quad ,
\eeq 
with
\bea
\hat{M}_2^-(d) &=& g_s \, (-i) f^{ABC} T^B T^C \,
 \mu^{2\epsilon} \frac{e^{\epsilon \gamma_E}}{(4 \pi)^{\epsilon}} \,
\frac 1 i \int \frac{d^d r}{(2 \pi)^d}
\bar{u}(p') \, \left( T^-_{\a \b} \, \varepsilon^\a \right)
\,  \frac{\psl + \rsl +m_b }{r^2+2(pr)} \, \g^\b \, u(p) \frac{1}{r^2}
\nonumber \\
\hat{M}_2^+(d) &=& g_s  \,\frac 3 2 T^A \,
\mu^{2\epsilon} \frac{e^{\epsilon \gamma_E}}{(4 \pi)^{\epsilon}}  \,
\frac 1 i \int \frac{d^d r}{(2 \pi)^d}
\bar{u}(p') \,  \left( T^+_{\a \b} \, \varepsilon^\a \right)
\,  \frac{\psl + \rsl +m_b }{r^2+2(pr)} \, \g^\b \, u(p) \frac{1}{r^2} \quad ,
\eea
where $T^+_{\a \b}$ and $T^-_{\a \b}$ are given in eqs. 
(\ref{tp}) and (\ref{tm}), respectively.
As the calculation of $\hat{M}_2^+(d)$ is  nothing but a repetition of 
the $b \to s \gamma$ case, we concentrate on $\hat{M}_2^-(d)$ in the following.
All the $\Delta i$ quantities in $T^-_{\a \b}$ contain the factor
$C^{-1-\e}$, whose explicit form is given in eq. (\ref{cs}).
$\hat{M}_2^-(d)$ can be written in  the form
\bea
\hat{M}_2^-(d) & = &\frac{g_s^3}{32\pi^2} \,(-i) f^{ABC} T^B T^C
\nn \\ & & \qquad
\mu^{2\epsilon} \frac{e^{\epsilon \gamma_E}}{(4 \pi)^{\epsilon}}  \,
\frac 1 i \int \frac{d^d r}{(2 \pi)^d}
\bar{u}(p') \, P(r) \, u(p) \,
\frac{[-\exp(i \pi \e)][x(1-x)]^{-1-\e}}{D_1 \, D_2 \,
D_3^{1+\e}} \quad ,
\eea
with
$D_1=(r^2+2(pr))$, $D_2=r^2$, $D_3= r^2 + 2(qr)y/(1-x) -m_c^2/(x(1-x))$.
The symbol
$P(r)$ is a matrix in Dirac space, which depends in a polynomial way
on the integration variable $r$.
In the next step, the three propagators $D_1$, $D_2$ and $D_3$
in the denominator
are Feynman parametrized as
\beq
\label{feynman}
\frac{1}{D_1 D_2 D_3^{1+\e}} = \frac{\G(3+\e)}{\G(1+\e)} \,
\int_S \frac{du dw  \, w^{\e}}{
\left[ r^2+2(pr)u +2(qr)yw/(1-x)-m_c^2 w/(x(1-x)) + i \delta \right]^{3+\e}}
\eeq
with $0 \leq w \leq 1-u$ and $0 \leq u \leq 1$. 
Then the integral over the loop momentum $r$ is performed.
At this level,  a four dimensional integral over the Feynman parameters
$(x,y;u,w)$  remains.
It is useful for the following
to perform the substitutions
\beq
\label{subst}
x \to x' ; \quad y \to -\frac{(1-x')(1-w'-y')}{w'} ; \quad
u \to (1-w') u' ; \quad w \to u' w'.
\eeq
The new variables then run in the intervals
\beq
 x',u',w' \in [0,1] ; \quad y \in [1-w',1].
\eeq
Taking into account the corresponding Jacobian
and omitting the primes ($'$) of the integration variables,
$\hat{M}_2^-(d)$ can be cast into the form
\bea
\label{m4ca}
\hat{M}_2^-(d) & = &
\frac{g_s^3}{32 \pi^2} \,(-i)f^{ABC}T^B T^C 
\nn \\ & & \qquad
\int \, dx dy du dw \, 
 \bar{u}(p') \,
\left[ F_1 \frac{\bar C}{\bar C^{2\e}} + F_2 \frac{1}{\bar C^{2\e}}
+ F_3 \frac{1}{\bar C^{1+2\e}} \right] \, u(p) \quad ,
\eea
where $F_1$, $F_2$ and $F_3$ are matrices in Dirac space
depending on the
Feynman parameters $x$, $y$, $u$, $w$. Note that this expression
is understood to be written in such a way that $F_1$, $F_2$ and $F_3$ are independent
of $m_c$. The charm quark mass then only enters through 
$\bar C$, which reads
\beq
\label{chat}
\bar C = m_b^2 u y  (1-w)  + \frac{m_c^2}{x(1-x)} w
\quad .
\eeq
In what follows, the ultraviolet $\e$ regulator remains a fixed,
small positive number.

The central point of our procedure
is to use now the Mellin-Barnes representation of the denominators that look
like propagators ($1/(k^2 -M^2)^\l$) \cite{Mellin},
which is given by ($\l>0$)
\beq
\label{Mellin}
\frac{1}{(k^2 - M^2)^\l} = \frac{1}{(k^2)^\l} \,
\frac{1}{\G(\l)} \, \frac{1}{2 \pi i} \, \int_{\gamma} ds
(-M^2/k^2)^s \G(-s) \G(\l+s) \quad .
\eeq
The symbol
$\gamma$ denotes the integration path which is parallel to the
imaginary axis (in the complex $s$-plane) hitting the real axis
somewhere between $-\l$ and $0$. In this formula,
the "momentum squared" $k^2$ is understood
to have a small positive imaginary part.

In our approach,
we use  formula (\ref{Mellin}) in order to simplify the
remaining Feynman parameter integrals in eq.
(\ref{m4ca}) where  we represent
 the factors
$1/\bar{C}^{2\e}$ and $1/\bar{C}^{1+2\e}$ 
as Mellin-Barnes integrals using the identifications
\beq
\label{ident}
k^2 \leftrightarrow m_b^2 u y (1-w)  \quad ; \quad
M^2 \leftrightarrow \frac{-m_c^2 \, w}{x(1-x)}  \quad .
\eeq
By interchanging the order of integration,
we first carry out the integrals over the Feynman parameters
for any given fixed value of $s$ on the integration path $\g$.
These integrals are basically the same as for the massless
case $m_c=0$ (in eqs. (\ref{m4ca}) and (\ref{chat})) up to
the factor 
\beq
\label{factor}
\left[ \frac{w}{ u \, y \, (1-w) \, x(1-x)} \right]^s
 \, \left( \frac{m_c^2}{m_b^2} \right)^s
\quad.
\eeq
Note that the functions $F_1$, $F_2$ and $F_3$
are such  that the Feynman parameter integrals exist
if the integration path $\g$ is properly chosen. In the terms
involving $F_2$ and $F_3$ in eq. (\ref{m4ca}),
the path must be chosen such that
$-\e < \Re(s) <0$; in the terms involving $F_1$ the situation is slightly
more complicated: $\bar{C}$ in the numerator should be replaced by the 
r.h.s of eq. (\ref{chat}). For the terms proportional to
$m_b^2$ the path has to be chosen as for the $F_2$ and $F_3$ contributions.
The terms proportional to $m_c^2$, however, lead to Feynman parameter
integrals which do not converge for values of $s$
on this path. It turns out that the
path has to be chosen such that $-2 \e < \Re(s) < -\e$ in order to have
convergent integrals for these terms.   

We would like to mention that the variable substitutions in eq. 
(\ref{subst}) were constructed in such a way that all the Feynman parameter
integrals are either elementary or of the form 
$\int_0^1 \, dx \, x^p (1-x)^q = \b(p+1,q+1)$. 

For the $s$ integration we use the residue
theorem after
closing the integration path
in the right $s$-halfplane. According to the above discussion,
the residue at $s=-\e$ has to be taken into account in the 
terms proportional to $m_c^2$.
In the other terms, however, the residue at $s=-\e$ must not 
be taken into account. The other poles
inside the
integration contour are
located at
\bea
s &=& 0, 1, 2, 3, ....... \nonumber \\
s &=& 1-\e, 2-\e, 3-\e, ....... \nonumber \\
s &=& 1-2\e, 2-2\e, 3-2\e, ....... \nonumber \\
s &=& 1/2-2\e, 3/2-2\e, 5/2-2\e, ....... \nonumber \\
s &=& 1-3\e, 2-3\e, 3-3\e, ....... \quad .
\eea

The other two-loop
diagrams are evaluated similarly.
The non-trivial Feynman integrals can always be reduced to
$\b$-functions after suitable substitutions.

The sum over the residues naturally leads to an
expansion in $z=(m_c^2/m_b^2)$ through the factor
$(m_c^2/m_b^2)^s$
in eq. (\ref{factor}).
This expansion, however, is not a Taylor series;
it also involves logarithms of $z$, which are generated
by the expansion in $\e$.
A generic diagram
which we denote by $G$ has then the form
\beq
\label{generic}
G = c_0 + \sum_{n,m} c_{nm} z^n
\ln^m z \quad ,  \quad z = \frac{m_c^2}{m_b^2} \quad .
\eeq
The power $n$ in eq. (\ref{generic})
is in general a natural multiple of $1/2$
and $m$ is a natural number including 0. In the explicit
calculation, the lowest $n$ turns out to be $n=1$.
This implies the important fact
 that the limit $m_c \to 0$ exists.

{}From the structure of the poles one can see that the power
$m$ of the logarithm is bounded by  $4$,
independent of the value of $n$. For a detailed explanation,
we refer to  \cite{GHW}.
As in this reference, we retain all terms up to $n=3$ in our
results.

Unlike in $b \to s \g$, the diagrams in the individual figures
are not gauge invariant. This statement holds even for the 
sum of all the diagrams in a) -- f) in fig. \ref{fig:1new}. A gauge invariant
result is only obtained after including the diagrams in g) and h)\footnote{We
 thank M. Neubert for making us aware of these diagrams.}.
We would like to mention that the diagrams analogous to g)
also exist for $b \to s \g$. Their sum, however, vanishes
in this case. As there are no gauge invariant subsets, we only
present the result which is obtained by summing all diagrams a) -- h) in fig.
\ref{fig:1new}. 
The result for $\hat{M}_2=\bra s g|\hat{O}_2|b\ket$
reads (using $z=(m_c/m_b)^2$ and $L=\ln z$):
\bea
\label{eq:virtualO2}
  \hat{M}_2    & = & \frac{1}{2592} \frac{\alpha_s}{\pi} \OEightTree  \left(
                \frac{m_b}{\mu} \right)^{-4 \epsilon} \nn \\
      &   &  \big\{ - \frac{384}{\epsilon} -2170 - 54 \pi^2 +z [48816 - 252
             \pi^2 + (22680 - 1620 \pi^2) L \nn \\
      &   &  \quad + 2808 L^2 + 612 L^3 -6480 \zeta(3)] \nn \\
      &   & \quad -12672 \pi^2 z^{3/2}+z^2 [66339 + 1872 \pi^2 + (-40446 +
            1512 \pi^2) L \nn \\
      &   & \quad + 6642 L^2 - 1008 L^3 + 7776 \zeta(3)] \nn  \\ 
      &   & \quad +z^3 [-3420 - 60 \pi^2 - 6456 L + 7884 L^2] \nn \\
      &   & \quad + 24 \pi i \big[ -28+z (549 - 24 \pi^2 + 153 L + 72
            L^2) \nn \\ 
      &   & \quad +z^2 (-432 + 30 \pi^2 + 54 L - 90 L^2)+z^3 (-259 +
            192 L) \big] \big\} \, .
\eea
In this expression, the symbol $\zeta$ denotes the Riemann
Zeta function, with $\zeta(3) \approx 1.2021$;
The symbol $\OEightTree$ denotes the tree level
matrix element of the operator $O_8$.
As such, it contains the 
 running $b$-quark mass and the running strong coupling constant,
both evaluated at the scale $\mu$ (see eq. (\ref{opbasis})).
However, as the corrections to $O_2$ are explicitly proportional
to $\a_s$, we are allowed (modulo higher order terms) to identify
the running $b$-quark mass with the pole mass $m_b$; in the same spirit
we can identify the strong coupling constant with $g_s(m_b)$. 
With this interpretation, which we will use in the
following, $\OEightTree$ is a scale independent quantity, reading
\beq
\label{o8tree}
\OEightTree = m_b \, \frac{g_s(m_b)}{8\p^2} \,
\bar{u}(p') \, \epsl \qsl R \,T^A \, u(p) \quad .
\eeq

We now turn to the matrix elements of the operator $\hat{O}_1$.
Due to the specific color structure it is straightforward to
see that only the diagrams e) and f) in fig. \ref{fig:1new} yield a
non-vanishing contribution, which is generated by the color symmetric
part of the building block $J_{\a \b}$ in eq. (\ref{build2}). The complete
regularized result for $\hat{M}_1=\bra s g|\hat{O}_1|b\ket$ reads
\bea
  \label{eq:O1summed}
  \hat{M}_1 & = &  \frac{1}{96} \frac{\alpha_s}{\pi} \OEightTree  \left(
                \frac{m_b}{\mu} \right)^{-4 \epsilon} \nn  \\
      &   & \big\{ - \frac{18}{\epsilon} -87+z [120 - 16 \pi^2 + (120 - 36
                \pi^2) L \nn \\
      &   & \quad  + 12 L^2 + 4 L^3 - 144 \zeta(3)] \nn \\
      &   & \quad +z^2 [84 + 32 \pi^2 - 24 \pi^2 L \nn \\
      &   & \quad - 12 L^2 + 
            4 L^3]+z^3 [-56 - 12 \pi^2 + 96 L - 36 L^2] \nn \\
      &   & \quad  - 4 \pi i \big[ 3+z (-24 + 2 \pi^2 - 6 L
             - 6 L^2)  \nn \\
      &   &  \quad +z^2 (-6 + 2 \pi^2 + 12 L - 6 L^2)-12 z^3 \big]
                \big\} \, .
\eea
The regularized matrix elements $M_1$ and $M_2$ of $O_1$ and $O_2$ 
in the operator basis (\ref{opbasis}) are related to $\hat{M}_1$
in eq. (\ref{eq:O1summed})
and $\hat{M}_2$ in eq. (\ref{eq:virtualO2}) as follows: 
\beq
\label{m1shift}
M_1 = \tfrac{1}{2} \, \hat{M}_1 - \tfrac{1}{6} \, \hat{M}_2 \ ; 
       \quad M_2 = \hat{M}_2 \,.
\eeq  

\subsection{Counterterms to the $\boldsymbol{O_1}$ and 
$\boldsymbol{O_2}$ contributions}
The operators mix under renormalization and thus the
counterterm
contributions must be taken into account. As we are 
interested in this
section in contributions to $b \to s g$ which are proportional to
$C_1$ and $C_2$, we have to include, 
in addition to the two-loop matrix elements
of $C_1 O_1$ and $C_2 O_2$, 
also the one-loop matrix elements of the four Fermi
operators
$C_i \delta Z_{ij} O_j$ ($i=1,2;j=1,...,6$) and the tree 
level contribution of
the magnetic operator $C_i \delta Z_{i8} O_8$ ($i=1,2$).
In the NDR scheme  the
only non-vanishing contributions
to $b \to s g$ come from  $j=4,8$ only. 
The operator renormalization
constants $Z_{ij}$ are  obtained from the leading
order anomalous dimension matrix
in the literature \cite{Misiak97} \footnote{Note that the
effective anomalous dimension matrix $\g^{0,\rm{eff}}$ given
in \cite{Misiak97} has to be converted into $\g^{0}$, before the relevant
$\d Z$-factors can be read off.}. 
The entries needed in our calculation
are
\beq
\label{zfactors1}
\delta Z_{14} =  -\frac{\a_s}{36 \pi \e}  \quad , \quad
\delta Z_{18} =   \frac{167 \, \a_s}{2592 \pi \e}  \quad .
\eeq
\beq
\label{zfactors2}
\delta Z_{24} =  \frac{\a_s}{6 \pi \e}  \quad , \quad
\delta Z_{28} =   \frac{19 \, \a_s}{108 \pi \e}  \quad .
\eeq
The counterterm contributions $M_1^{\rm{ct}}$ and $M_2^{\rm{ct}}$
proportional to $C_1$ and $C_2$ are then given by
\beq
\label{ct1}
M_1^{\rm{ct}} = \bra s g |\delta Z_{14} O_4+\delta Z_{18} O_8|b \ket
= \left( \frac{\a_s}{216 \pi} \,
  \frac{1}{\e} \left(
\frac{m_b}{\mu} \right)^{-2\e} +
\frac{\a_s}{ \pi} \, \frac{167}{2592}
\, \frac{1}{\e} \right) \OEightTree \quad .
\eeq
\beq
\label{ct2}
M_2^{\rm{ct}} = \bra s g |\delta Z_{24} O_4+\delta Z_{28} O_8|b \ket
= \left( -\frac{\a_s}{36 \pi} \,
  \frac{1}{\e} \left(
\frac{m_b}{\mu} \right)^{-2\e} +
\frac{\a_s}{ \pi} \, \frac{19}{108}
\, \frac{1}{\e} \right) \OEightTree \quad .
\eeq

We note that there are no one-loop contributions to 
the matrix element for $b \to s g$ from the counterterms
proportional to the evanescent operators $P_{11}$ and 
$P_{12}$ given in
appendix A of ref. \cite{Misiak97}.
\subsection{Renormalized matrix elements of $\boldsymbol{O_1}$ and $\boldsymbol{O_2}$}
Adding the regularized two-loop result in eq. (\ref{eq:virtualO2})
and the counterterm in
eq. (\ref{ct2}), we find the renormalized result
for $M_2$ in the NDR scheme:
\beq
\label{m2lr}
M_2^{\rm{ren}} = \OEightTree \, \frac{\a_s}{4 \p} \,
\left( \ell_2 \ln \frac{m_b}{\mu}  + r_2 \right) \quad ,
\eeq
with
\beq
\label{l2}
\ell_2 = \frac{70}{27}
\eeq
\bea
\label{rer2ndr}
\Re( r_2 ) & = &  \frac{1}{648}
                          \big\{ -2170 - 54 \pi^2 + z [48816 - 252 \pi^2 + (22680
                         - 1620 \pi^2) L \nonumber \\
                   &   & \quad + 2808 L^2 + 612 L^3 - 6480 \zeta(3)] \nonumber
                          \\
                   &   & \quad -12672 \pi^2 z^{3/2}+z^2 [66339 + 1872 \pi^2 +
                          (-40446 + 1512 \pi^2) L \nonumber \\ 
                   &   & \quad + 6642 L^2 - 1008 L^3 + 7776
                          \zeta(3)] \nonumber \\
                   &   & \quad +z^3 [-3420 - 60 \pi^2 - 6456 L + 7884
                          L^2] \big\} \nn
\eea
\bea
\label{imr2ndr}
\Im( r_2 ) & = &  \frac{\p}{27}
                       \big\{
                  -28+z [549 - 24 \pi^2 + 153 L + 72
                         L^2] \nonumber \\ 
                   &   & \quad +z^2 [-432 + 30 \pi^2 + 54 L - 90 L^2]+z^3 [-259 +
                         192 L]  \big\}
\eea
Here, $\Re(r_2)$ and $\Im(r_2)$ denote the real and the imaginary part
of $r_2$, respectively. The quantity $z$ is defined as $z=(m_c^2/m_b^2)$
and $L=\ln(z)$.
\begin{figure}[t]
\begin{center}
\leavevmode
\includegraphics[scale=1]{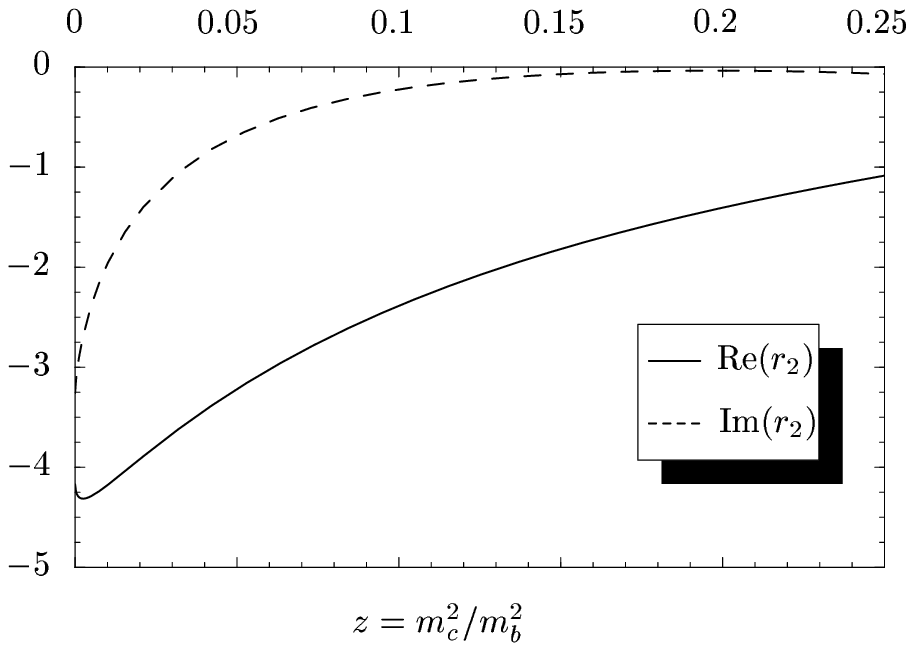}
\end{center}
\caption[f1]{
Real and imaginary part of $r_2$ in the NDR scheme
(from eq.
(\ref{rer2ndr})).}
\label{r2figure}
\end{figure}
\begin{figure}[t]
\begin{center}
\leavevmode
\includegraphics[scale=1]{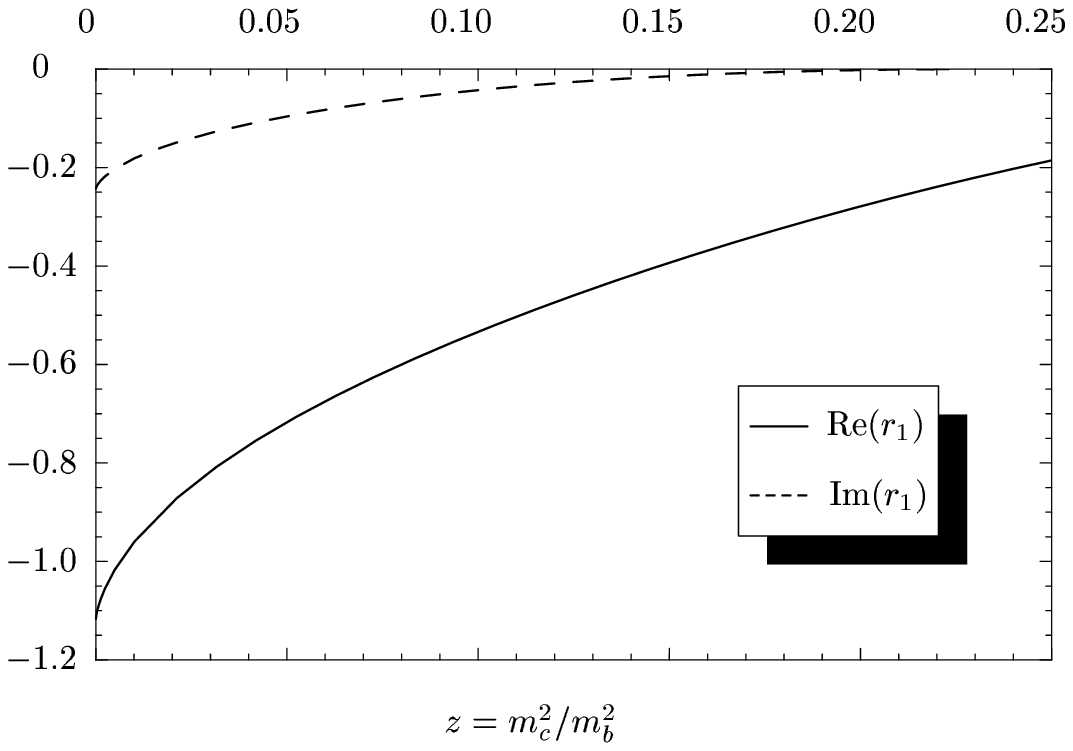}
\end{center}
\caption[f1]{
Real and imaginary part of $r_1$ in the NDR scheme
(from eq.
(\ref{rer1ndr})).}
\label{r1figure}
\end{figure}

Similarly, we obtain the renormalized version of $M_1$ by 
adding the regularized two-loop result in eq. (\ref{m1shift})
and the counterterm in
eq. (\ref{ct1}); we find 
\beq
\label{m1lr}
M_1^{\rm{ren}} = \OEightTree \, \frac{\a_s}{4 \p} \,
\left( \ell_1 \ln \frac{m_b}{\mu}  + r_1 \right) \quad ,
\eeq
with
\beq
\label{l1}
\ell_1 = \frac{173}{162}
\eeq
\bea
\label{rer1ndr}
\Re( r_1 )  & = &  -\frac{1}{3888}
                          \big\{ 4877 - 54 \pi^2 + 36 z [1086 + 29 \pi^2 + (360
                         + 36 \pi^2) L \nonumber \\
                   &   & \quad + 51 L^2 + 8 L^3 + 144 \zeta(3)] \nonumber
                          \\
                   &   & \quad -12672 \pi^2 z^{3/2}+ 9 z^2 [6615 - 80 \pi^2 +
                          (-4494 + 384 \pi^2) L \nonumber \\ 
                   &   & \quad + 864 L^2 - 148 L^3 + 864
                          \zeta(3)] \nonumber \\
                   &   & \quad +12 z^3 [93 + 76 \pi^2 - 1186 L + 900
                          L^2] \big\} \nn
\eea
\bea
\label{imr1ndr}
\Im( r_1 ) & = &  -\frac{\pi}{324}
                       \big\{
                       25 +6 z [75 + \pi^2 + 24 L -3  L^2] \nonumber \\ 
         &   & \quad +6 z^2 [-171 + 19 \pi^2 + 72 L - 57 L^2]+ 2 z^3 [-421 +
                         192 L]  \big\}
\eea
In figs. \ref{r2figure} and \ref{r1figure}
we show the real and the imaginary parts of
$r_2$ and $r_1$, respectively.
For  $z \ge 1/4$ the imaginary parts must
vanish exactly; indeed  we see
from these plots that the imaginary parts
based on the expansion retaining terms up to
$z^3$
indeed
vanish at $z=1/4$ to high accuracy.
\section{Virtual corrections to $\boldsymbol{O_8}$}
\label{virto8}
In this section we calculate the order $\a_s$ virtual 
corrections to the matrix element
\beq
\label{mato8}
M_8 = \bra s g |O_8|b\ket \ .
\eeq
%
%\begin{figure}[t]
%\begin{center}
%\leavevmode
%\epsfxsize=6.0 truecm
%\epsfbox[209 322 502 452]{O83abc.ps}
%\end{center}
%\caption[f1]{Diagrams associated with the operator $O_8$.}
%\label{figo8:1}
%\end{figure}
%
\begin{figure}[t]
\begin{center}
\leavevmode
\includegraphics[scale=1]{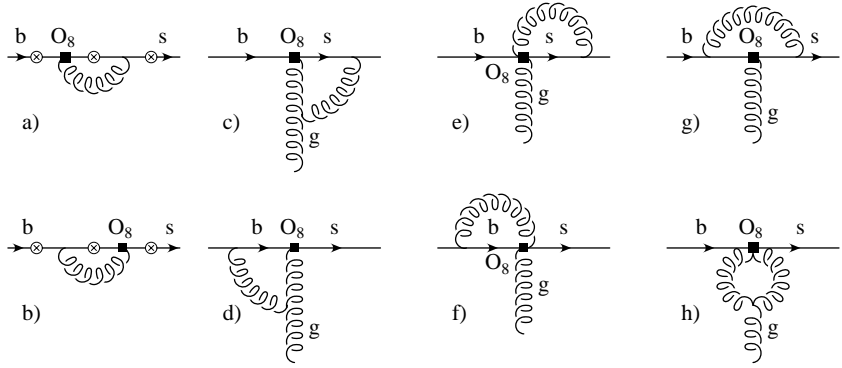}
\vspace{3ex}
\caption[f1]{Diagrams associated with the operator $O_8$. The real gluon can
  be attached to any of the circle-crosses on the fermion lines.}
\label{fig:o8virtual}
\end{center}
\end{figure}
As the contributing Feynman graphs in fig. 
\ref{fig:o8virtual} are one loop diagrams,
the computation of $M_8$ is  straightforward.
We use dimensional regularization for both, the ultraviolet and 
the infrared singularities. Singularities which appear in the situation
where the virtual gluon becomes almost real and collinear with the
emitted gluon are also regulated dimensionally; on the other hand,
those singularities where the almost real internal gluon
is collinear with the $s$-quark, are
regulated with a small strange quark mass $m_s$;
the latter manifest themselves in logarithmic terms
of the form $\ln(\rho)$, where $\rho=(m_s/m_b)^2$.
 
We were able to separate the ultraviolet $1/\e$ poles from those
which are of infrared (and/or collinear) origin. For ultraviolet poles
we use the symbol $1/\e$ in the following, while collinear and infrared
poles are denoted by $1/\e_{\rm IR}$.

When working in Feynman gauge for the gluon propagator, the individual
diagrams contributing to $M_8$ have the following infrared/collinear
properties (the letters refer to the diagrams in fig. \ref{fig:o8virtual}): 
a) and b)  are free of infrared and collinear singularities;
c) has combined infrared/collinear singularities of the form $1/\e^2_{\rm IR}$
   or  $\ln(\rho)/\e_{\rm IR}$ as well as $1/\e_{\rm IR}$ poles;
d) has combined infrared/collinear singularities of the form $1/\e^2_{\rm IR}$
   as well as $1/\e_{\rm IR}$ poles;
e) has a collinear singularity of the form $\ln(\rho)$;
f) is free of infrared and collinear singularities;
g) has a combined collinear and infrared singularity of the form
   $\ln(\rho)/\epsilon_{\rm IR}$ as well as collinear singularities of the form
   $\ln^2(\rho)$ and $\ln(\rho)$;
h) has an infrared singularity of the form
   $1/\e_{\rm IR}$; more precisely, this diagram is proportional to the
   combination $(1/\e-1/\e_{\rm IR})$.

As the results of the individual diagrams are not very instructive,
we only give their sum:
\beq
\label{o8result}
M_8 = \frac{\a_s}{4\p} \, f_8 \,  \OEightTree \ ,
\eeq
with
\bea
\label{f8}
f_8 &=& \left[ \,- \frac{3}{\e_{\rm IR}^2} - \frac{(4 \ln(\rho)+9+9i\p)}{3\e_{\rm IR}} +
 \frac{11}{3\e} \right] \, \left( \frac{m_b}{\mu} \right)^{-2\e} + 
 \nn \\
& & \quad \frac{1}{3} \, \left[ \frac{59\p^2}{12} + 1
     -8 \ln(\rho) + 2 \ln^2(\rho)-8i\p \right] \ .
\eea

We would like to mention that we did not include diagrams with
self energy insertions in the external legs. As we work in an on-shell
renormalization scheme with respect to quark and gluon fields, such diagrams
are cancelled against counterterm contributions.

\subsection{Counterterms to the $\boldsymbol{O_8}$ contribution}
The counterterm is generated by expressing the bare
quantities in the tree-level matrix element of $O_8$
by their renormalized counterparts. It has the structure
\beq
\label{m8ct}
M_8^{\rm{ct}} = \delta R \, \OEightTree \quad ,
\eeq
where the factor $\delta R$ is given by
\beq
\label{Rfactor}
\delta R = \sqrt{Z_2(m_b)} \, \sqrt{Z_2(m_s)} \, \sqrt{Z_3} \, Z_{g_s} \,
Z_{m_b} \, Z_{88} \, - \,  1 \ .
\eeq
$Z_2(m_b)$, $Z_2(m_s)$ and $Z_3$ denote the on-shell 
wave function renormalization factors
of the $b$-quark, the $s$-quark and the gluon, respectively.
$Z_{g_s}$ and $Z_{m_b}$ denote the $\overline{\mbox{MS}}$ 
renormalization constants for
the strong coupling constant $g_s$ and the $b$-quark mass factor, which
appear explicitly in the definition of the operators (see eq.
(\ref{opbasis})).  
Finally, $Z_{88}$ is the renormalization factor of the operator $O_8$.

The explicit form of $Z_2(m)$ reads
\beq
\label{Z2}
Z_2(m) = 1 - \frac{\a_s}{3\p}  \, \left( \frac{m}{\m} 
 \right)^{-2\e} \, \left[ \frac{1}{\e} + \frac{2}{\e_{\rm IR}} +4 \right] \,
\eeq
where we again separated infrared and ultraviolet poles.
For $Z_3$ we get in the on-shell scheme:
\beq
\label{Z3}
Z_3 = 1 + \frac{\a_s}{2\p} \frac{5}{2} \, \left( \frac{1}{\e} 
- \frac{1}{\e_{\rm IR}} \right) - \frac{\a_s}{2\p} \frac{1}{3} \sum_f
\left[ \frac{1}{\e} - 2 \ln \frac{m_f}{\m} \right] \ .
\eeq
The sum in this formula run over the five flavors $f=u,d,c,s,b$.
For $Z_{m_b}$ and $Z_{88}$ (see ref. \cite{Misiak97}) one obtains
\beq
Z_{m_b} = 1 - \frac{\a_s}{4\p} \, \frac{4}{\e} \ ; \quad
Z_{88} = 1 +  \frac{\a_s}{4\p} \, \frac{14}{3\e} \ .
\eeq
Finally, the renormalization constant for the strong coupling constant
reads
\beq
Z_{g_s} = 1 -  \frac{\a_s}{4\p} \, \left[ \frac{11}{2} - \frac{N_f}{3}
\right] \, \frac{1}{\e} \ ; \quad N_f=5 \ .
\eeq 
Inserting the various $Z$ factors in eq. (\ref{Rfactor}), one
obtains
\beq
\label{Rexplizit}
\delta R = -\frac{\a_s}{4\p} \, \left[
\frac{11}{3\e} + \frac{31}{6\e_{\rm IR}} - 8 \ln \frac{m_b}{\m}
-\frac{2}{3} \sum_f \ln \frac{m_f}{\m} + \frac{16}{3} - 2 \ln\rho
\right]\ .
\eeq

\subsection{Renormalized matrix element of $\boldsymbol{O_8}$}
Adding the regularized matrix element of $O_8$ in eq. 
(\ref{o8result}) and
the counterterm contribution $M_8^{\rm{ct}}$ in eq. (\ref{m8ct}),
one obtains the renormalized result
\beq
\label{m8ren}
M_8^{\rm{ren}} = \frac{\a_s}{4\p} \, f_8^{\rm{ren}} \, \OEightTree \ ,
\eeq
with
\bea
\label{f8ren}
f_8^{\rm{ren}} &=& \left[ \, - \frac{3}{\e_{\rm IR}^2} - 
     \frac{(8 \ln(\rho)+49+18i\p)}{6\e_{\rm IR}} \right] \,
  \left( \frac{m_b}{\mu} \right)^{-2\e} -
\frac{29}{3} \ln \frac{m_b}{\m} \nn \\
& & \quad +\frac{2}{3} \sum_f \ln \frac{m_f}{\m} \,
     -5 + \frac{59\p^2}{36} -\frac{2}{3} \ln\rho+ \frac{2}{3} \ln^2 \rho
     - \frac{8}{3} i \p \ .
\eea
In eq. (\ref{f8ren}) the sum runs over the five flavors $f=u,d,c,s,b$,
and $\rho=(m_s/m_b)^2$.
We anticipate that the singular terms of the form $1/\e^2_{\rm IR}$, $1/\e_{\rm IR}$ 
and $\ln \rho$ in eq. (\ref{f8ren}) will cancel 
(at the level of the decay width) 
against the corresponding singularities 
present in the gluon bremsstrahlung corrections to $b \to s g$. On the
other hand, the logarithmic terms
$\ln(m_f/\m)$, which also represent some kind of singularities
for the light flavor $f=u,d,s$ are not cancelled by the gluon bremsstrahlung
process. Keeping in mind that these terms originate from the renormalization 
factor $Z_3$ of the gluon field, i.e.,
from gluon self energy 
diagrams in which these flavors propagate, 
it is expected that these logarithms will cancel against
the logarithms present in the decay rate $\G(b \to s f \bar{f})$ with 
$f=u,d,s$. To cancel these unphysical terms, we will  include 
the $O_8$ contribution 
to this process in section \ref{Quarkstrahlung}.
   
\section{Virtual corrections to the decay width for
$\boldsymbol{\symbol{98} \to \symbol{115} \symbol{103} }$}
\label{Gammavirt}
We are now ready to write down the renormalized version of 
the matrix $M^{\rm{ren}}(b \to s g)$ 
element for $b \to s g$, where the virtual order
$\a_s$ corrections are included. We obtain:
\bea
\label{matelvirt}
M^{\rm{ren}}(b \to s g) &=& \frac{4G_F i }{\sqrt{2}} \, V_{ts}^* V_{tb} \,
\bigg\{ C_8^{\rm{eff}}  + 
\frac{\a_s}{4\p} \bigg[ C_1^0 (\ell_1 \ln \frac{m_b}{\mu}+r_1) +
                        C_2^0 (\ell_2 \ln \frac{m_b}{\mu}+r_2) +
 \nn \\
& &  \quad
       C_8^{0,\rm{eff}} \, f_8^{\rm{ren}} \, \bigg]
\bigg\} \, \bra sg|O_8(\mu)|b \ket_{\rm{tree}}  .
\eea
The quantities $\ell_1$, $r_1$, $\ell_2$, $r_2$, and $f_8^{\rm{ren}}$ 
are given in eqs. (\ref{l1}), (\ref{imr1ndr}), (\ref{l2}), 
(\ref{imr2ndr}) and (\ref{f8ren}), respectively. 
As eq. 
(\ref{matelvirt}) shows, 
$C_8^{\rm eff}$ is the only Wilson coefficient needed to NLL precision.
For the following, it is useful to decompose it as
\beq
\label{c8decomp}
C_8^{\rm{eff}} = C_8^{0,\rm{eff}} + \frac{\a_s}{4\p} \, C_8^{1,\rm{eff}} \ . 
\eeq
The symbol
$\bra s g|O_8(\mu)|b \ket_{\rm{tree}}$ in eq. (\ref{matelvirt})
denotes the tree level matrix
element of $O_8(\mu)$, which contains the running $b$-quarks mass
and the strong running coupling constant at the scale $\mu$.  
In order to get expressions where the $b$- quark
mass enters as the pole mass, and the strong coupling constant enters
as $g_s(m_b)$, we rewrite       
$\bra s g|O_8(\mu)|b \ket_{\rm{tree}}$ as
\beq
\label{treerel}
\bra s g|O_8(\mu)|b \ket_{\rm{tree}} =
\bra s g|O_8|b \ket_{\rm{tree}} \, \left[
1 + \frac{2\a_s}{\p} \ln \frac{m_b}{\m} - \frac{4}{3} \frac{\a_s}{\p}
+\frac{\a_s}{4\p} \, \b_0 \, \ln \frac{m_b}{\m} \right] \ ; \quad
\b_0=\frac{23}{3}  \ ,
\eeq
where we made use of eqs. (\ref{polerunning}) and (\ref{aqcd}).
The symbol $\bra s g|O_8|b \ket_{\rm{tree}}$ then stands for
the tree level matrix element of $O_8$ in which $\overline{m}_b(\m)$
and $g_s$ have to
to be identified with the pole mass $m_b$ and $g_s(m_b)$, respectively.
(See also the discussion after eq. (\ref{eq:virtualO2}) and eq. 
(\ref{o8tree})). 
Inserting eqs. (\ref{c8decomp}) and (\ref{treerel}) into eq. 
(\ref{matelvirt}) we obtain: 
\bea
\label{matelvirtnew}
M^{\rm{ren}}(b \to s g) &=& \frac{4G_F i }{\sqrt{2}} \, V_{ts}^* V_{tb} \,
\bigg\{
 C_8^{0,\rm{eff}}  + 
\frac{\a_s}{4\p} \bigg[ C_8^{1,\rm{eff}} + 
(8 +\b_0) \, \ln \frac{m_b}{\m} \, C_8^{0,\rm{eff}}  
                        -\frac{16}{3} \, C_8^{0,\rm{eff}} +
                \nn \\ 
       && \quad
   C_1^0 (\ell_1 \ln \frac{m_b}{\mu}+r_1)+ 
                        C_2^0 (\ell_2 \ln \frac{m_b}{\mu}+r_2) +
                        C_8^{0,\rm{eff}} \, f_8^{\rm{ren}} \, \bigg]
\bigg\} \, \bra sg|O_8|b \ket_{\rm{tree}} \ .
\eea
To obtain the decay width $\G^{\rm{virt}}$ 
from $M^{\rm{ren}}(b \to s g)$  
is straightforward. We get:
\bea
\label{gammavirt}
\G^{\rm{virt}} &=&  
 \frac{\a_s(m_b) \, m_b^5 }{24\p^4} |G_F V_{ts}^* V_{tb}|^2 \,
\left\{ \left( C_8^{0,\rm{eff}} \right)^2 + 
\frac{\a_s}{4\p} \, C_8^{0,\rm{eff}} \,\left[
2 \, C_8^{1,\rm{eff}} 
+2(8+\b_0) \ln \frac{m_b}{\m} \, C_8^{0,\rm{eff}} \right . \right. \nn \\
& &
-\frac{32}{3} C_8^{0,\rm{eff}} 
+ 2 \, C_1^0 
(\ell_1 \, \ln \frac{m_b}{\m} + \Re(r_1)) +
2 \, C_2^0 (\ell_2 \, \ln \frac{m_b}{\m} +\Re(r_2))  \nn \\ 
& & \left. \left. \quad + 2 \, C_8^{0,\rm{eff}} \Re(f_8^{\rm{ren}}) \, (1-\e)
\, \left( \frac{m_b}{\mu} \right)^{-2\e} \,
\left( 1+2\e-\frac{1}{4}(\pi^2-16)\e^2 \right) \right] \right\}  \ .
\eea
We note that due to the infrared poles present in $f_8^{\rm{ren}}$ 
the phase space integrations have been done consistently in $d=4-2\e$
dimensions, which leads to the last two extra factors in the last term 
in eq. (\ref{gammavirt}). The other factor, $(1-\e)$, in the
last term in eq. (\ref{gammavirt}), is due to the fact that
all the $(d-2)$ possible transverse polarizations of the emitted gluon
were taken into account.

\section{$\boldsymbol{O_8}$ contribution to the decay width 
 $\boldsymbol{\G(\symbol{98} \to \symbol{115} \symbol{102} \bar{\symbol{102}}})$}
\label{Quarkstrahlung}
As discussed at the end of section \ref{virto8}, we should take into 
account the contribution of the operator $O_8$ to the process
$b \to s f \bar{f}$ $(f=u,d,s)$, in order to cancel the unphysical logarithms
of the form $\ln(m_f/\m)$ in the virtual corrections to $b \to s g$.
The $O_8$ contribution to the 
decay width $\G_8(b \to s f \bar{f})$ yields
\beq
\label{quarkdecay}
\G_8(b \to s f \bar{f}) = \frac{m_b^5 \, |G_F \, V_{ts}^*V_{tb} 
\, C_8^{0,\rm{eff}}|^2}{72\p^5} \, \a_s^2 \, \left[ \ln \frac{m_b}{2m_f} 
-\frac{2}{3} \right] \ .
\eeq 
Comparing this result with $\G^{\rm{virt}}$ in eq. (\ref{gammavirt}),
we see explicitly, that the mentioned logarithms indeed cancel.
  
\section{Matrix elements for gluon bremsstrahlung}
\label{bremsmat}
In this section we discuss the gluon bremsstrahlung corrections 
to $b \to s g$, i.e. the matrix element for the process 
$b \to s g g$, associated with the
operators $\hat{O}_1$, $\hat{O}_2$ and $O_8$. 
For literature on the analogous corrections to $b \to s \gamma$, we refer
to \cite{ALI}. 
\subsection{Bremsstrahlung associated with $\boldsymbol{\hat{O}_1}$ and 
$\boldsymbol{\hat{O}_2}$}
We first discuss the matrix
element 
of $\hat{O}_2$. 
There are two diagrams contributing; they are displayed in d) and e)
of fig. \ref{fig:o8o2brems}. The sum of diagram d) and the one with
the two gluons interchanged is denoted by
%The first diagram (actually it's the sum of diagram d) plus the diagram where
%the two gluons are exchanged), denoted by 
$\bar{J}_{\a\b}$. Its analytic form
is obtained by putting $r^2=0$ and $r_\b=0$ in the expression for 
$J_{\a \b}$
in eq. (\ref{build2}): 
\beq
\label{brems2}
\bar{J}_{\a \b}^{AB} = \bar{T}^+_{\a \b}(q,r) \left\{T^A , 
T^B \right\} +
                 \bar{T}^-_{\a \b}(q,r) \left[T^A ,T^B 
\right] \quad . 
\eeq
This expression is understood to be contracted with the polarization
vectors $\varepsilon^\a(q)$ and $\varepsilon^\b(r)$ of the gluons.
The diagram e) in fig. \ref{fig:o8o2brems}, denoted by $S^{AB}_{\a\b}$,
is color antisymmetric and can be written as 
\beq
S^{AB}_{\a\b} = S^-_{\a\b} \, 
       \left[T^A ,T^B \right] \quad , 
\eeq 
where $S^-_{\a\b}$ reads ($t=(2qr)/m_c^2$)
\beq
S^-_{\a\b} = \frac{g_s^2}{32\p^2} \, \left[ \frac{4}{3} \left( 
\frac{\m}{m_c} \right)^{2\e} 
\frac{1}{\e} - \frac{4}{3} - 8 G_1(t) + 8 G_2(t) \right]
 \, \left[ \rsl g_{\a\b} 
 - \qsl g_{\a\b} 
- 2 \g_\b r_\a +2 \g_\a q_\b \right] \, L \ .
\eeq
The functions $G_i(t)$ ($i=-1,0,1,...$) are defined as
\beq
\label{gi}
G_{i}(t) = \int_0^1 dx \, x^i \, \ln [ 1-tx(1-x)-i \delta] \quad .
\eeq
The Ward identities $r^\b \bar{T}^+_{\a \b}=q^\a \bar{T}^+_{\a \b}=0$,
stated in \cite{GHW}, imply that  
\beq
\label{tpbrems}
\bar{T}^+_{\a \b}  = \frac{g^2_s}{32 \pi^2} \,
\left[  E(\a,\b,r) 
       - E(\a,\b,q) 
       - E(\b,r,q) \, \frac{r_\a}{(qr)} 
       + E(\a,r,q) \, \frac{q_\b}{(qr)} 
       \right] \, L  \, \bar\Delta i_{23} \quad .
\eeq
General considerations (or a straightforward calculation which makes use
of the explicit expressions for the functions $G_i$ and $\bar{\D}i_i$)
imply the Ward identities
\beq
r^\b (\bar{T}^-_{\a\b} + S^-_{\a\b})=0 \ ; \quad 
q^\a (\bar{T}^-_{\a\b} + S^-_{\a\b})=0 \ ,
\eeq
which can be used to cast $(\bar{T}^-_{\a\b}+S^-_{\a\b})$ into 
the simple form
\beq
\label{tmsmbrems}
\bar{T}^-_{\a\b} + S^-_{\a\b} = \frac{g_s^2}{32\p^2} \, (\rsl -\qsl) \,
\left( \frac{r_\a q_\b}{qr} - g_{\a\b} \right) \, L \, \bar{\Delta}i_{17}
\ .
\eeq

To summarize, the matrix element $\hat{M}_2^{\rm{brems}}=
\bra sgg|\hat{O_2}|b\ket$ can be written as
\beq
\label{m2brems}
\hat{M}_2^{\rm{brems}} = 
\bar{T}^+_{\a \b} \left\{ T^A , 
T^B \right\} + \left( \bar{T}^-_{\a \b} + S^-_{\a \b} \right)
 \left[T^A ,T^B \right] \quad , 
\eeq
where $\bar{T}^+_{\a \b}$ and 
      $(\bar{T}^-_{\a \b} + S^-_{\a\b})$ are given in eqs. (\ref{tpbrems}) 
and (\ref{tmsmbrems}), respectively. The functions 
$\bar{\Delta}i_{23}$ and $\bar{\Delta}i_{17}$ occurring in these
expressions, can be written in terms of $G_0(t)$ and $G_{-1}(t)$
($t=(2qr)/m_c^2$) defined in eq. (\ref{gi}):
\beq
\label{deltabar}
\bar{\Delta}i_{23} = -2 \, \frac{t+2\, G_{-1}(t)}{t} \ ; \quad
\bar{\Delta}i_{17} = -\frac{2}{3} \, \frac{t+6\, G_{-1}(t) -12 \, G_0(t)}{t}
 \ .
\eeq
The explicit form of $G_{-1}(t)$ and $G_0(t)$ is given in appendix
\ref{appendix:b}.
Note that these results are ultraviolet finite. As the subsequent
phase space integrals do not generate infrared singularities,
it is consistent to retain  terms up to order 
$\e^0$ only in eq. (\ref{m2brems}).

Due to the specific color structure of the operator $\hat{O}_1$,
the  diagram e) in fig. \ref{fig:o8o2brems} does not contribute and the
color antisymmetric part encoded in $\bar{T}^-_{\a \b}$ is also absent.
The matrix element $\hat{M}_1^{\rm{brems}}=
\bra sgg|\hat{O_1}|b\ket$ is therefore proportional to $\bar{T}^+_{\a \b}$,
reading
\beq
\label{m1brems}
\hat{M}_1^{\rm{brems}} = 
\bar{T}^+_{\a \b} \, \d^{AB} \, \d^{ab} \ ;
\eeq
$A,B$ and $a,b$ are the color indices of the gluons and the quarks, 
respectively.
\subsection{Bremsstrahlung associated with $\boldsymbol{O_8}$}
\begin{figure}[t]
\begin{center}
\leavevmode
\includegraphics[scale=1]{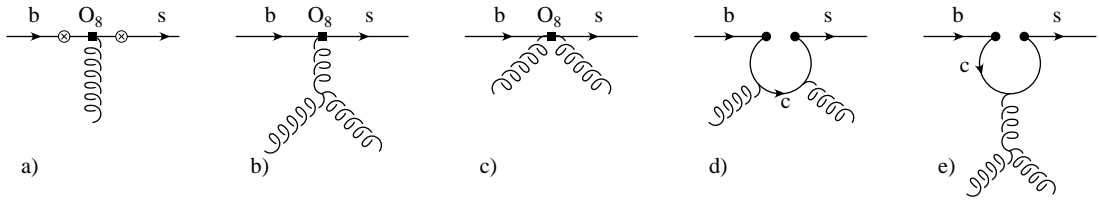}
\end{center}
\caption[f1]{Bremsstrahlung diagrams associated to $O_8$ and
  $\hat{O}_2$. Circle-crosses denote possible gluon emissions. Note that 
  picture a) actually represents four Feynman diagrams (obtained
by interchanging the gluons) 
and the one in d) represents two diagrams (again:
  including the interchange of the gluons).}
\label{fig:o8o2brems}
\end{figure}

The Feynman diagrams contributing to 
the matrix element $M_8^{\rm{brems}}=\bra s g g |O_8| b \ket$ 
are shown in a), b) and c) in fig. \ref{fig:o8o2brems}.
Similar to $\hat{M}_2^{\rm{brems}}$ in eq. (\ref{m2brems}),
one can decompose $M_8^{\rm{brems}}$ into a color symmetric- and
a color antisymmetric part:
\beq
\label{m8brems}
M_8^{\rm{brems}} = 
R^+_{\a \b} \left\{ T^A , T^B\right\} + R^-_{\a \b} 
 \left[T^A ,T^B \right] \quad . 
\eeq
The diagrams shown in b) and c)
only contribute to
$R^-_{\a\b}$, while the diagrams in a)
contribute to both
$R^-_{\a\b}$ and $R^+_{\a\b}$. As the calculation of these tree level
diagrams is straightforward, we do not give the explicit expressions
for $R^+_{\a\b}$ and $R^-_{\a\b}$.

\section{Decay width for $\boldsymbol{\symbol{98} \to \symbol{115} \symbol{103} \symbol{103}}$}
\label{bremswidth}
The total matrix element $M^{\rm{brems}}(b \to s g g)$  
can be written as
\beq
\label{mbrems}
M^{\rm{brems}} = \frac{4G_F i}{\sqrt{2}} \, V_{ts}^* V_{tb} \, \left[
\hat{C}_1 \, \hat{M}_1^{\rm{brems}} +  \hat{C}_2 \,\hat{M}_2^{\rm{brems}} +
C_8^{0,\rm{eff}} \, M_8^{\rm{brems}} \right] \ ,
\eeq
where the three terms on the r.h.s., given in eqs. (\ref{m1brems}),
(\ref{m2brems}) and (\ref{m8brems}), correspond to the contributions
of the operators $\hat{O}_1$,  $\hat{O}_2$ and $O_8$, respectively.
The coefficients $\hat{C}_1$ and $\hat{C}_2$ are understood to be
the following linear
combinations of the Wilson coefficients $C_1$ and $C_2$ appearing in the
effective Hamiltonian (\ref{heff}):
\beq
\hat{C}_1 = \tfrac{1}{2} \, C_1 \ ; \quad
\hat{C}_2 = C_2 - \tfrac{1}{6} \, C_1 \ ; \quad
\eeq
We note that in eq. (\ref{mbrems}) only the leading order pieces of the
Wilson coefficients are needed.

\noindent
The expression for the decay width reads in $d$ dimensions:
\beq
\label{dgamma}
d \G^{\rm{brems}}(b \to s gg) = 
\frac{1}{2m_b} \, \int \, (2\p)^d  \d^d(p-p'-q-r) \, 
\overline{|M^{\rm{brems}}|^2_\Sigma} \,
d\m(p') d\m(q) d\m(r) \ ,
\eeq
where $p$, $p'$, $q$, $r$ are the four-momenta of the
$b$-quark, $s$-quark,  and the gluons. 
$\overline{|M^{\rm{brems}}|^2_\Sigma}$ is obtained by squaring
the matrix element
$M^{\rm{brems}}$, followed by summing/averaging  over spins and color 
of the final/initial state particles. The factor $(1/2)$ due to the
two gluons in the final state is also absorbed there. 

The phase space integrals are plagued with infrared and collinear 
singularities. Configurations with one gluon flying collinear to the
$s$-quark are regulated by a small strange quark mass $m_s$,
while 
configurations with two collinear gluons, or one soft gluon
are dimensionally regularized. 
As in the calculations of the virtual corrections, we
write the dimension as $d=4-2\e$. (Note that $\e$ has to be negative
in order to regulate the phase space integrals). 
%
%We would like to mention that we summed over all the $d-2$ transverse
%polarizations of the emitted gluon when calculating the virtual corrections
%to $b \to s g$. For consistency, we also sum over all the $d-2$ transverse
%polarizations of the two gluons in the bremsstrahlung process.

When squaring $M^{\rm{brems}}$ in eq. (\ref{mbrems}), nine terms
are generated, which we denote for obvious reasons 
by $(\hat{O}_1,\hat{O}_1^*)$, $(\hat{O}_1,\hat{O}_2^*)$,
$(\hat{O}_1,O_8^*)$, $(\hat{O}_2,\hat{O}_1^*)$, $(\hat{O}_2,\hat{O}_2^*)$,
 $(\hat{O}_2,O_8^*)$, $(O_8,\hat{O}_1^*)$, 
$(O_8,\hat{O}_2^*)$,
and $(O_8,O_8^*)$. It turns out that all terms except $(O_8,O_8^*)$
are free of infrared and collinear singularities. We therefore
can put $m_s=0$ in these terms and evaluate
the phase space integrals in $d=4$ dimensions. 
Denoting this finite contribution to the decay width by
$\G^{\rm{brems}}_{\rm{fin}}$, we get:
\beq
\label{bremsfin}
\G^{\rm{brems}}_{\rm{fin}} = \frac{8 \, |G_F V_{ts}^*V_{tb}|^2}{64 \p^3 
\, m_b} \, \frac{1}{12} \, \frac{\alpha_s^2}{64 \pi^2} 
\, \int dE_q \, dE_r
          (\tau_{11}^+ + \tau_{22}^{+} + \tau_{22}^{-} +
      \tau_{12}^{+} +    \tau_{18}^+ + \tau_{28}^{+} + \tau_{28}^{-} ) \ .
\eeq
The superscripts $(+)$ and $(-)$ on the various $\tau$-quantities 
refer to color even and color odd contributions, respectively.
The result is represented 
as a two dimensional integral over the energies $E_q$ and
$E_r$ of the gluons in the rest frame of the $b$-quark.
$E_q$ and $E_r$ vary in the range
\beq
  E_q \in \left[0,\frac{m_b}{2}\right] \ ; 
\quad E_r \in \left[\frac{m_b}{2} - E_q,
  \frac{m_b}{2}\right] \ .
\eeq
The various $\tau$-quantities, in which all the scalar products are 
understood to be expressed in terms of $E_q$ and $E_r$, read:
\bea
\label{taus}
  \tau_{11}^{+}     & = & \hat{C}_1^2 \;  24   \; | \bar{\D} i_{23} |^2 \; 
                        2 m_b^2 [m_b^2 - 2
                      (q r)] \nn \\
  \tau_{22}^{+} & = & \hat{C}_2^2 \; \tfrac{28}{3} \; | \bar{\D} i_{23} |^2  \;
                      2 m_b^2 [m_b^2 - 2 (q r)] \nn \\
  \tau_{22}^{-} & = & \hat{C}_2^2 \;  12 \; | \bar{\D} i_{17} |^2 \; 2 \,
                      [16(pq)^2-16(pq)(qr)-8m_b^2(pq)+6m_b^2(qr)+m_b^4] \nn \\
  \tau_{12}^{+}     & = & 2 \, \hat{C}_1 \, \hat{C}_2 \;  8 
                         \; | \bar{\D} i_{23} |^2 \; 
                        2 m_b^2 [m_b^2 - 2
                      (q r)] \nn \\
  \tau_{18}^{+}     & = & 2 \hat{C}_1 \, C_8^{0,\rm{eff}} \;  8  \; 
                       \Re(\bar{\D} i_{23}) \; 16
                      m_b^2 (q r) \nn \\
  \tau_{28}^{+} & = & 2 \hat{C}_2 \, C_8^{0,\rm{eff}} \; \tfrac{28}{3} \;  
                       \Re(\bar{\D} i_{23}) \; 
                       16 m_b^2 (q r) \nn \\ 
  \tau_{28}^{-} & = & 2 \hat{C}_2 \, C_8^{0,\rm{eff}} \; 12 \;
                       \Re(\bar{\D} i_{17}) \, (-4m_b^2) \,
                       \left[ m_b^4 (p q) +  m_b^4 (p r)
                      - 2 m_b^2 (p q)^2 - 2 m_b^2 (p r)^2  \right. \nn \\
                &   & \left.  \quad \quad - 2 m_b^2 (p q) (p r) + 
                      4 (p q)^2 (p r) + 4 (p r)^2 (p q) \right]/
                      [(pq)(pr)]
\eea
were the functions  $\bar{\D} i_{17}$ and  $\bar{\D} i_{23}$ are given in 
eq. (\ref{deltabar}). As these function are rather complicated, the 
integrals over $E_q$ and $E_r$ are done numerically.

We now turn to the $(O_8,O_8^*)$ contribution, denoted 
by $\G^{\rm{brems}}_{88}$. 
Without going too much into the details,
we would like to mention that some care has to be taken 
when summing over the $(d-2)$ transverse polarizations of the gluons.
These sums are of the form
\beq
\label{polsum}
\sum_{r=1}^{d-2} \varepsilon^\m_r(k) \, \varepsilon^{*\n}_r(k) =
- g^{\m\n} + k^\m f^\n + k^\n f^\m \, ,
\eeq 
where the vector $f$ satisfies the condition $(fk)=1$,
with $k$ being the four-momentum of the gluon. It turns
out that both terms involving $f$ on 
the r.h.s in eq. (\ref{polsum}) contribute to
the color antisymmetric part of $\G^{\rm{brems}}_{88}$.
After a lengthy, but straightforward calculation, 
we obtain (with $\rho=(m_s/m_b)^2$)
\beq
\label{brems88p}
\G^{\rm{brems},+}_{\rm{88}} = 
\frac{7  \a_s \left( C_8^{0,\rm{eff}} \right)^2 \, V}{96 \, \p} \,
 \left( \frac{m_b}{\m} \right)^{-4\e}
\, \left[ \frac{8 + 4 \ln \rho}{\e_{\rm IR}} - 
2 \ln^2\rho+6 \ln\rho+ 18 - \frac{4\p^2}{3} 
\right] 
\eeq
for the color symmetric part, and
\beq
\label{brems88m}
\G^{\rm{brems},-}_{\rm{88}} = 
\frac{\a_s \left( C_8^{0,\rm{eff}} \right)^2 \, V}{16 \, \p} \,
\, \left( \frac{m_b}{\m} \right)^{-4\e}
\, \left[ \frac{24}{\e^2_{\rm IR}} + \frac{80 + 6 \ln \rho}{\e_{\rm IR}} -
3 \ln^2\rho+9 \ln\rho+299 -26\p^2  
\right]
\eeq
for the color antisymmetric part. $V$ is defined as
\beq
 V = \frac{\a_s \, m_b^5}{24\p^4} \, |G_F V_{ts}^*V_{tb}|^2 \, .
\eeq
The total decay with for $b \to s g g$ is then 
given by
\beq
\label{gammabremstot}
\G^{\rm{brems}}(b \to s g g ) =
\G^{\rm{brems}}_{\rm{fin}}+
\G^{\rm{brems},+}_{88}+
\G^{\rm{brems},-}_{88} \ ,
\eeq
where the three terms on the r.h.s. are given in eqs. (\ref{bremsfin}),
(\ref{brems88p}) and (\ref{brems88m}).
%
%%%%%%%%%%%%%%%%%%%%%%%%%%%%%%%%%%%%%%%%%%%%%%%%%%%%%
\section{Combined NLL branching ratio for $\boldsymbol{\symbol{98} \to
    \symbol{115}\symbol{103}}$ and $\boldsymbol{\symbol{98} \to \symbol{115}\symbol{103}\symbol{103}}$}
\label{brnll}
In this section we combine the decay widths for the virtually
corrected process
$b \to s g$ and the bremsstrahlung process $b \to s g g$ to the decay width,
which we call $\G^{\rm{NLL}}(b \to s g)$. 
We also absorb in this quantity the $O_8$ induced contribution to the
process $b \to s f \bar{f}$, where $f=u,d,s$, as discussed at the end
of section \ref{virto8} and in section \ref{Quarkstrahlung}. The expression
for $\Gamma^{\rm{virt}}$, which contains the lowest order contribution
to the decay width for $b \to s g$, together with its virtual corrections,
may be found in eq. (\ref{gammavirt}). The result for the bremsstrahlung
process, $\Gamma^{\rm{brems}}$ is given in eq. (\ref{gammabremstot}).
{}From the explicit formulas for $\Gamma^{\rm{virt}}$ and $\Gamma^{\rm{brems}}$
one can see that the infrared singularities and those 
collinear singularities, which are regulated by 
$\e_{\rm IR}$ cancel in the sum. The same also happens with the collinear
singularities which are regularized by the parameter $\rho=(m_s/m_b)^2$.
The terms containing logarithms of the light quark masses $m_f$,
present in the result for $\Gamma^{\rm{virt}}$, are cancelled when combined
with  $\G_8(b \to s f \bar{f})$ in eq. (\ref{quarkdecay}). 
Putting together the individual pieces, we obtain
\bea
\label{gammanll}
\G^{\rm{NLL}}(b \to s g) &=&  
 \frac{\a_s(m_b) \, m_b^5 }{24\p^4} |G_F V_{ts}^* V_{tb}|^2 \,
\left\{ \left( C_8^{0,\rm{eff}} \right)^2 + 
\frac{\a_s}{4\p} \, C_8^{0,\rm{eff}} \,\left[
2 \, C_8^{1,\rm{eff}} 
-\frac{32}{3} C_8^{0,\rm{eff}} \right . \right. \nn \\
& & + 2 \, C_1^0 
[\ell_1 \, \ln \frac{m_b}{\m} + \Re( r_1)] +
2 \, C_2^0 [\ell_2 \, \ln \frac{m_b}{\m} +\Re( r_2)]  \nn \\ 
& & \left. \left. \quad + 2 \, 
C_8^{0,\rm{eff}} [(\ell_8+8+\b_0) \, \ln \frac{m_b}{\m} + r_8]  
\right] \right\} + \G_{\rm{fin}}^{\rm{brems}} \ ,
\eea
where $\G_{\rm{fin}}^{\rm{brems}}$, given in eq. (\ref{bremsfin}), 
contains all the bremsstrahlung corrections except those originating
from the $(O_8,O_8^*)$ interference. The quantities $\ell_1$, $r_1$, 
$\ell_2$ and $r_2$ stem from the virtual corrections; they are given
in eqs. (\ref{l1}), (\ref{rer1ndr}), (\ref{l2}) and (\ref{rer2ndr}), 
respectively. On the other hand, $\ell_8$ and $r_8$ contain information from
the real part of the virtual corrections, encoded in $\Re( f_8^{\rm{ren}})$;
the contributions from the $(O_8,O_8^*)$ interference of the gluon
bremsstrahlung process; and the $O_8$ contribution to the process
$b \to s f \bar{f}$:
The explicit expressions for $\ell_8$ and $r_8$ (which is real by definition) 
read
\beq
\label{l8r8}
\ell_8=-\frac{19}{3} \ ; \quad
r_8=\frac{1}{18} \, \left[ 351 -19 \p^2 -36 \ln2 + 
6 \ln \frac{m_c^2}{m_b^2} \right] \ .
\eeq   
We would like to stress that all scale dependent quantities in eq. 
(\ref{gammanll}) are understood to be evaluated at the scale $\mu$, unless
indicated explicitly in the notation.

To prepare the discussion on the numerical size of the NLL QCD corrections,
it is useful to cast the final result (\ref{gammanll}) into another form:
\beq
\label{rewrite}
\G^{\rm{NLL}}(b \to s g) =  
\frac{\a_s(m_b) \, m_b^5 }{24\p^4} |G_F V_{ts}^* V_{tb}|^2 \,
|\bar{D}|^2 + \G_{\rm{fin}}^{\rm{brems}} \ ,
\eeq
with
\bea
\label{dbar}
\bar{D}  &=& C_8^{0,\rm{eff}} + 
\frac{\a_s}{4\p} \, \left[
C_8^{1,\rm{eff}} 
- \frac{16}{3} C_8^{0,\rm{eff}}
+ C_1^0 [\ell_1 \, \ln \frac{m_b}{\m} + r_1]  \right. \nn \\
& & \left. \quad \quad \quad
+ C_2^0 [\ell_2 \, \ln \frac{m_b}{\m} +r_2]
+ C_8^{0,\rm{eff}} [(\ell_8+8+\b_0) \, \ln \frac{m_b}{\m} + r_8] \right]  \ .
\eea
The modulus square of $\bar{D}$ is understood to be taken in the same
way as the in the virtual contributions, i.e., by systematically
discarding the $O(\a_s^2)$ term. In this sense, the quantity
$\bar{D}$ can be viewed as an effective matrix element.

We would like to mention that $\ell_1$, $\ell_2$ and $(\ell_8+8+\b_0)$
are identical to the anomalous dimension matrix elements
$\g^{0,\rm{eff}}_{18}$,  
$\g^{0,\rm{eff}}_{28}$, and  $\g^{0,\rm{eff}}_{88}$, respectively.
This is of course what has to happen: Only in this case the
leading scale dependence of $C_8^{0,\rm{eff}}(\mu)$ gets compensated
by the second term in eq. (\ref{dbar}).

The NNL branching ratio ${\cal B}^{\rm{NLL}}(b \to s g)$ 
is then obtained as
\beq
\label{BRdef}
{\cal B}^{\rm{NLL}}(b \to s g) = \frac{\G^{\rm NLL}(b \to s g)}{\G_{\rm{sl}}} \,
{\cal B}_{\rm{sl}}^{\rm{exp}} \ ,
\eeq
where ${\cal B}_{\rm{sl}}^{\rm{exp}}$ denotes the experimental
semileptonic branching ratio of the $B$-meson. $\G_{\rm{sl}}$
stands for the theoretical expression of the semileptonic decay width of the
$B$-meson. Neglecting non-perturbative corrections of the order 
$(\L_{\rm{QCD}}/m_b)^2$, $\G_{\rm{sl}}$ reads (with $x_c=(m_c/m_b)$)
\beq
\label{gammasl}
\Gamma_{\rm{sl}} \approx \G(b \to c e \bar{\n}_e) =
\frac{G_F^2 \, m_b^5}{192\p^3} \, |V_{cb}|^2 \, g(x_c) \,
\left[1 + \frac{\a_s(\mu_b)}{2\p} \, h_{\rm{sl}}(x_c) + O(\a_s^2) \right] \ ,
\eeq
where the phase space function $g(x_c)$ reads
\beq
g(x_c) = 1 - 8 \, x_c^2 - 24 \, x_c^4 \, \ln x_c + 8 \, x_c^6 - x_c^8 \ .
\eeq
The analytic expression for $h_{\rm{sl}}(x_c)$ can be found in ref.
\cite{Nir}. The approximation
\beq
h_{\rm{sl}}(x_c)=-3.341 + 4.05 \, (x_c-0.3) - 4.3 \, (x_c-0.3)^2
\eeq
holds to an accuracy of 1 permille in the relevant 
range $0.2 \le x_c \le 0.4$.

We note that in the numerical analysis of ${\cal B}^{\rm{NLL}}(b \to s g)$
we systematically expand 
the expression for the branching ratio (\ref{BRdef}) in $\a_s$, dropping
terms of $O(\a_s^2)$.

A short remark concerning the LL branching ratio is in order: For the
decay width $\G^{\rm{LL}}(b \to s g)$, we use the expression
\beq
\G^{\rm{LL}}(b \to s g) =  
\frac{\a_s(\mu) \, m_b^5 }{24\p^4} |G_F V_{ts}^* V_{tb}|^2 \,
\left( C_8^{\rm LL, eff}(\mu)\right)^2 \ .
\eeq
The LL branching ratio for $b \to s g$ is then obtained as in eq. 
(\ref{BRdef}), but by discarding the radiative corrections in 
$\G_{\rm{sl}}$.  

\section{Numerical results for the combined branching ratio}
\label{numres}
\begin{figure}[t]
\begin{center}
\leavevmode
\includegraphics[scale=0.5]{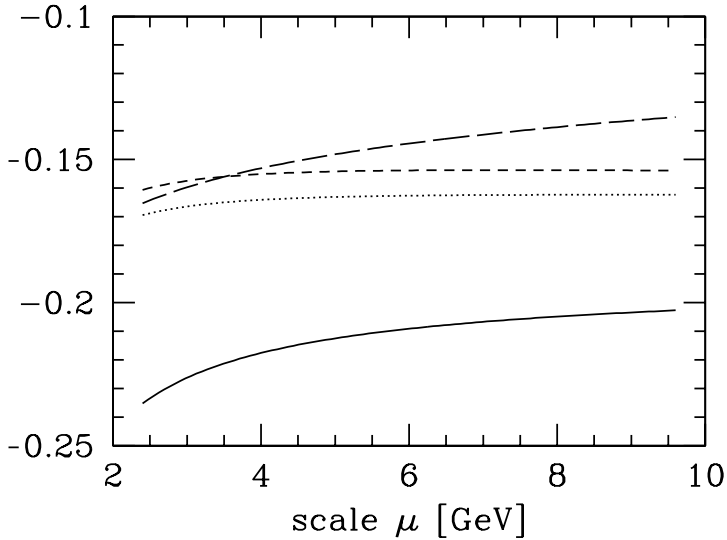}
\end{center}
\caption[f1]{Scale ($\mu$) dependence of the function $\bar{D}$
(see eq. (\ref{dbar})) in various approximations: The long-dashed line shows
$C_8^{0,\rm{eff}}$; the short-dashed line corresponds to putting
$r_1=r_2=r_8=0$; the dotted line is obtained by only putting $r_2=0$;
the solid line shows the full function $\bar{D}$. See text.}
\label{fig:dfun}
\end{figure}

Before we present the numerical result for the branching ratio
${\cal B}^{\rm{NLL}}(b \to s g)$, we discuss the sizes of the various
NLL corrections at the level of the function $\bar{D}$, defined in
eq. (\ref{dbar}) (anticipating that the finite bremsstrahlung corrections
in eq. (\ref{rewrite}) are relatively small).
We already mentioned that the terms containing the explicit
logarithms of the ratio $(m_b/\m)$ get compensated by the scale dependence
of the first term on the r.h.s. of eq. (\ref{dbar}). This feature
can be observed in fig. \ref{fig:dfun}, when comparing the two dashed
lines. The long-dashed line represents only the first term $C_8^{0}$
of the function $\bar{D}$, while the short-dashed
line shows $\bar{D}$, in which  $r_1$, $r_2$ and $r_8$
are put to zero. As expected, the short-dashed line has a milder 
$\mu$-dependence. When switching on also $r_1$ and $r_8$ 
(but keeping $r_2=0$), the resulting curve, shown by the dotted line,
stays close to the short-dashed curve and the scale dependence remains
mild. However, when switching on also $r_2$, the situation changes
drastically. The resulting solid line, which  represents the full NLL 
$\bar{D}$ function, implies that the term containing
the two-loop quantity $r_2$, induces a large NLL correction.
As this large correction term contains a factor $\a_s(\mu)\, C_2(\mu)$,  
it is of no surprise, that the NLL prediction for the function 
$\bar{D}$ suffers from a relatively large scale dependence, as illustrated
by the solid line.

\begin{figure}[t]
\begin{center}
\leavevmode
\includegraphics[scale=1]{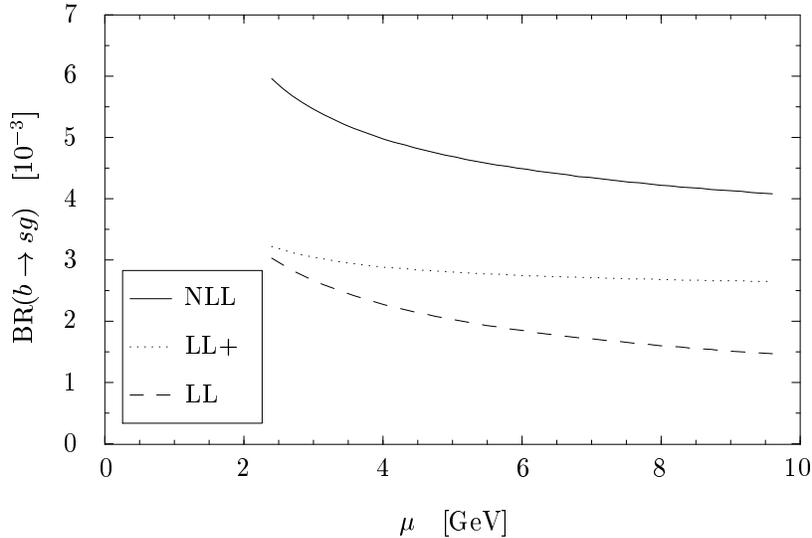}
\end{center}
\caption[f1]{Branching ratio ${\cal B}(b \to s g)$ as a function
of the scale $\mu$ in various approximations: The dashed and the solid
lines show the LL and the NLL predictions, respectively; the dotted
line is obtained by putting $r_1=r_2=r_8=\G_{\rm{brems}}^{\rm{fin}}=0$
in the NLL expression for $\G^{\rm{NLL}}(b \to s g)$ in 
eq. (\ref{rewrite}). See text.}
\label{fig:br}
\end{figure}

The NLL branching ratio ${\cal B}^{\rm{NLL}}(b \to s g)$ is then obtained
as described in section \ref{brnll}. The result is shown by the solid
line in fig. \ref{fig:br}. For the input values, we take:
$m_b=(4.8 \pm 0.2)$ GeV, $(m_c/m_b)=0.29 \pm 0.02$, 
$\a_s(m_Z)=0.119 \pm 0.003$, 
$|V_{ts}^* \, V_{tb}/V_{cb}|^2=0.95 \pm 0.03$,
${\cal B}_{\rm{sl}}^{\rm{exp}}=(10.49 \pm
0.46)\%$, and $m_t^{\rm{pole}}=(175 \pm 5)$ GeV.
As the scale dependence is rather large, we did not take
into account the error due to the uncertainties in the input parameters. 
Based on fig. \ref{fig:br}, we obtain the NLL branching ratio
\beq
\label{brvalue_nll}
{\cal B}^{\rm{NLL}}(b \to s g) = (5.0 \pm 1.0) \times 10^{-3} .
\eeq
We would like to stress that the NLL corrections drastically enhance
the LL value (see dashed line in fig. \ref{fig:br}) for which one obtains
\beq
\label{brvalue_ll}
{\cal B}^{\rm{LL}}(b \to s g) = (2.2 \pm 0.8) \times 10^{-3} \ .
\eeq
As already mentioned in the discussion of the function $\bar{D}$,
the main enhancement is due to the virtual- and bremsstrahlung
corrections to $b \to s g$, calculated in this paper.
At the level of the branching ratio, this fact is illustrated by 
the dotted line in fig. \ref{fig:br}, 
which is obtained by discarding $\G_{\rm{fin}}^{\rm{brems}}$
and by switching off $r_1$, $r_2$ and $r_8$ in  the expression
for $\G^{\rm{NLL}}(b \to s g)$ (see eq. (\ref{rewrite})).

\begin{figure}[t]
\begin{center}
\leavevmode
\includegraphics[scale=1]{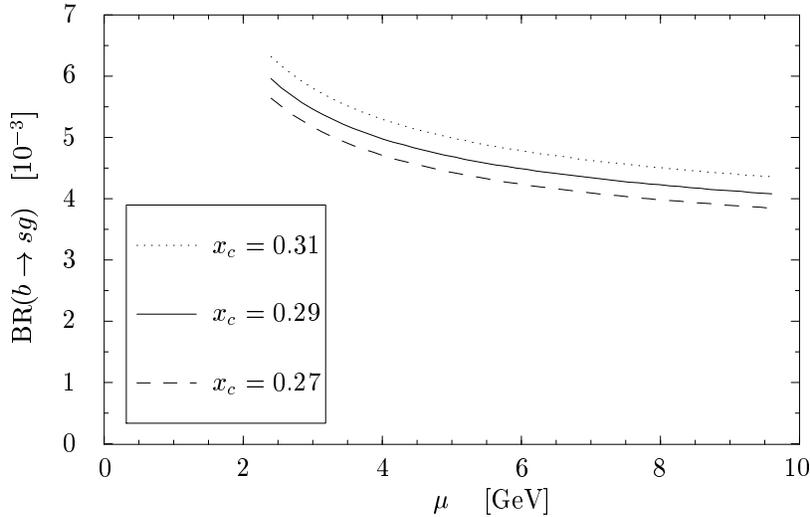}
\end{center}
\caption[f1]{NLL Branching ratio ${\cal B^{\rm{NLL}}}(b \to s g)$ as a function
of the scale $\mu$ for the three value of the ratio $x_c=m_c/m_b$. See text.}
\label{fig:brmc}
\end{figure}

The largest uncertainty due to the physical input parameters 
on ${\cal B}^{\rm{NLL}}(b \to s g)$ results from the charm quark mass.
The dependence of ${\cal B}^{\rm{NLL}}(b \to s g)$ on $m_c$ is illustrated
in fig. \ref{fig:brmc}, where $x_c=m_c/m_b$ is varied between 0.27 and
0.31. Choosing $\mu=m_b$, the resulting uncertainty amounts to
$\sim \pm 6 \%$.  
\section{Numerical evaluation of the charmless decay rate}
\label{charmless}
In this section we investigate the impact of the NLL QCD corrections
to $b \to s g$ on the inclusive hadronic charmless decay rate of 
the $\overline{B}$ meson.
At the quark level, we take into account 
%the semileptonic decays
%\beq
%\label{semi}
%b \to u \ell^- \bar{\nu}_\ell \ ; \quad \ell = \mu,e 
%\eeq
%and 
the hadronic processes 
\beq
\label{had}
b \to q' \bar{q}' q \ ; \quad b \to s g \, ,
\eeq
where $q=d,s$ and $q'=u,d,s$. As we do not distinguish between
$\Delta S=0$ and
$\Delta S=1$ contributions, we can safely neglect the CKM suppressed
decay mode $b \to d g$.
%As usual we normalize the corresponding decay widths by the 
%semileptonic decay width $\G_{\rm{sl}}$ (see eq. (\ref{gammasl})).
More precisely, we calculate the CP-averaged branching ratio
\beq
\label{brc}
\brc =  \frac{\Gamma(b \to \xc) + 
\Gamma(\overline{b} \to \barxc)}{2 \, \G_{\rm{sl}}} \, 
{\cal B}_{\rm{sl}}^{\rm{exp}} \, ,
\eeq
where $\xc$ stands for the final states listed in eq. 
(\ref{had}). In the numerical results for $\brc$ we will 
insert $\G_{\rm{sl}}$ as given in eq. (\ref{gammasl}), i.e.,
we do not make an $\a_s$ expansion of $1/\G_{\rm{sl}}$ 
in eq. (\ref{brc}). 
The charmless hadronic decay rate $\brc$ then reads
%\beq
%\rc= \sum_{q=d,s;q'=u,d,s} \, r_{q'\bar{q}'q} + r_{sg} + 
% r_{ue} + r_{u\mu} ,
%\eeq
\begin{eqnarray}
\label{brcdecomp}
   \brc & = &  \brsg + 
                        % \CPBR{ue} + \CPBR{u\mu} +
   \sum_{ \scriptsize
     \begin{array}{l}
       q\phantom{'}  =  d,s \\ 
       q' =  u,d,s
     \end{array}}
   \brqqq \ .
\end{eqnarray}
%where e.g. $r_{u\bar{u}d}$ is defined as
%\beq
%r_{u\bar{u}d} = \frac{\G(b \to u \bar{u} d)}{\G_{\rm{sl}}} \ .
%\eeq
%
%We omit small contributions such as 
%%$r_{u \tau}=0.004$, $r_{d g}$ 
%$\CPBR{u \tau} \;[\Gamma(b \to u \tau)/\Gamma_{\rm sl}=0.004]$, $\CPBR{d g}$
%and radiative decay modes. 
%
While the $O(\a_s)$ corrections to semileptonic processes
have been known for a long time (see e.g. ref. \cite{Nir}), 
the NLL corrections to the hadronic processes
in eq. (\ref{had})
with 3 quarks in the final state had a long history and 
were completed to a large extent only recently by Lenz et al.
\cite{Lenz1,Lenz2}; 
however, current-current
type corrections to the penguin operators are still missing. 
To briefly summarize the history, it is useful
to decompose the NLL expressions for the decay widths of these processes
into various pieces. Taking as an example the process $b \to u \bar{u} d$,
we write as in ref. \cite{Lenz1}:
\beq
\label{gammapeng}
\G(b \to u \bar{u} d) = \G^{(0)} + \frac{\a_s}{4\p} \, \left[
\D\G_{\rm{cc}} +  \D\G_{\rm{peng}} + \D\G_8 +  \D\G_{\rm{W}} \right] \ .
\eeq

The first two terms in the square bracket in eq. (\ref{gammapeng}) describe
the effect of current-current and penguin diagrams involving
the operators\footnote{Note, that the authors of refs. \cite{Lenz1,Lenz2}
use the old operator basis \cite{Grinstein90}.}
$O_1$ and $O_2$. $\D\G_8$ likewise contains the matrix element of the
operator $O_8$. The remaining part $\D\G_{\rm{W}}$ of the NLL contribution
is made of the corrections to the Wilson coefficients multiplying the
tree-level amplitudes in $\G^{(0)}$. 
In this approximation, the matrix elements of the
penguin operators $O_3$,...,$O_6$ only enter at tree level.
As the expressions for the r.h.s. of eq. (\ref{gammapeng}) are explicitly
given in ref. \cite{Lenz1}, we do not give them here. 
For later reference, we denote this approximation 
(in lack of a better word) by ``approx1''.

Later, in ref. \cite{Lenz2}, the same authors added the contributions
of the penguin diagrams associated with the penguin operators to the 
decay matrix elements and took into account the interference
with the tree level matrix element of the operator $O_2$ in the 
decay width. In addition, they  took into account the square of the
matrix element of the penguin diagram
associated with $O_2$. Although being of next-to-NLL,
this term is numerically relatively large, as it is multiplied with $C_2^2$.
These new contributions can be absorbed into the quantity 
$\D \G_{\rm{new}}$, which is understood to be added to the terms
in the bracket in eq. (\ref{gammapeng}). As the extraction of 
$\D \G_{\rm{new}}$ from ref. \cite{Lenz2} is straightforward,
we do not give the explicit expression.
This approximation, which contains --  
up to the current-current
type corrections to the penguin operators --
the full NLL contribution
to the hadronic three body decays, is called ``approx2''.

We note that the approximation where only the current-current
type corrections $\D\G_{\rm{cc}}$ were considered together with the shifts
$\D\G_{\rm{W}}$ in the Wilson coefficients has existed for a
long time \cite{Altarelli}. We denote this approximation 
by ``approx0'' in the numerical discussion.

In table \ref{tab:BRNoCharm} we present numerical results for 
the charmless hadronic branching ratio $\brc$ in the various approximations
mentioned above. 
The process $b \to s g$, encoded in 
$\brsg$ in eq. (\ref{brcdecomp}) 
is taken into
account in the columns ``approx0'', ``approx1'' and``approx2'' at LL precision.
The last column includes in addition the NLL corrections to $b \to s g$
which were calculated in this paper.
\begin{table}[htbp]
  \begin{center}
    \begin{tabular}{l|r|r|r|r}
      input 
    & approx0 \hspace{5ex} & approx1 \hspace{5ex} & approx2 \hspace{5ex} &
      with NLL $b\to s g$  \hspace{5ex} \\   \hline \hline 
       as in (\ref{input})        & 1.32 & 1.50  & 1.62  & 1.88 \\ \hline
      $ \mu = m_b/4 $               & 3.86 & 3.21  & 3.34  & 3.62 \\ \hline
      $ \mu = m_b/2 $               & 2.06 & 2.09  & 2.18  & 2.43 \\ \hline
      $ \mu = 2 m_b $               & 0.96 & 1.14  & 1.28  & 1.55 \\ \hline \hline
      $ |V_{ub} / V_{cb}| = 0.06  $ & 0.94 & 1.13  & 1.24  & 1.50 \\ \hline
      $ |V_{ub} / V_{cb}| = 0.07  $ & 1.03 & 1.22  & 1.33  & 1.59 \\ \hline  
      $ |V_{ub} / V_{cb}| = 0.08  $ & 1.14 & 1.32  & 1.44  & 1.69 \\ \hline
      $ |V_{ub} / V_{cb}| = 0.09  $ & 1.26 & 1.44  & 1.55  & 1.81 \\ \hline  
      $ |V_{ub} / V_{cb}| = 0.10  $ & 1.39 & 1.57  & 1.69  & 1.94 \\ \hline
      $ |V_{ub} / V_{cb}| = 0.11  $ & 1.54 & 1.72  & 1.83  & 2.09 \\ \hline
      $ |V_{ub} / V_{cb}| = 0.12  $ & 1.70 & 1.87  & 1.99  & 2.25 \\ \hline
      $ |V_{ub} / V_{cb}| = 0.13  $ & 1.87 & 2.05  & 2.16  & 2.42 \\ \hline \hline
      $ x_c = 0.25 $                & 1.14 & 1.32  & 1.45  & 1.69 \\ \hline
      $ x_c = 0.27 $                & 1.22 & 1.41  & 1.53  & 1.78 \\ \hline
      $ x_c = 0.29 $                & 1.32 & 1.50  & 1.62  & 1.88 \\ \hline
      $ x_c = 0.31 $                & 1.44 & 1.61  & 1.72  & 1.99 \\ \hline
      $ x_c = 0.33 $                & 1.57 & 1.74  & 1.84  & 2.12 \\
          \end{tabular}
    \caption{Table for the charmless hadronic branching ratio
    $\brc$ (in $\%$) in the various approximations
    discussed in the text. Unless specified explicitly in the first column,
    the input parameters correspond to the central
    values in eq. (\ref{input}).} 
    \label{tab:BRNoCharm}
  \end{center} 
\end{table}
Table \ref{tab:BRNoCharm} was produced with the following input
parameters:
\bea
\label{input}
& &
m_b=(4.8 \pm 0.2) \mbox{ GeV}, \quad \mu=m_b, \quad (m_c/m_b)=0.29 \pm 0.04, 
\nn \\
& &
\a_s(m_Z) = 0.119 \pm 0.003, 
\quad m_t^{\rm{pole}}=(175 \pm 5) \mbox{ GeV}, \quad
{\cal B}_{\rm{sl}}^{\rm{exp}}=(10.49 \pm 0.46)\%
 \nn \\
& &
|V_{us}|=0.22, \quad |V_{cb}|=0.038, \quad 
|V_{ub}/V_{cb}|=0.095 \pm 0.035,
\quad \delta=60^o \pm 30^o .
\eea
The central value for $|V_{ub}/V_{cb}|$ corresponds to the (improved)
Wolfenstein parameters $\overline{\rho}=0.20$ and $\overline{\eta}=0.37$
\cite{Ali_London}. The remaining entries of the CKM matrix are then obtained
as described in detail in \cite{Buras95}. We note that
the averaged charmless
hadronic branching ratio is practically independent of $\delta$, as
already observed in ref. \cite{Lenz2}.

The numbers in  column ``approx2'' are very similar to those in
table 1 of ref. \cite{Lenz2}. The small discrepancy is due to the omission
of the $1/m_b^2$ power corrections in our work.

Staring from the numbers in column ``approx0'', table \ref{tab:BRNoCharm}
illustrates, that the various improvements shown in the other columns
are relatively large, tending to increase $\brc$. In particular, the
NLL corrections to $b \to s g$ are of similar importance as the corrections
calculated in \cite{Lenz1,Lenz2}.

For $|V_{ub}/V_{cb}|=0.095$ we obtain the charmless hadronic 
branching ratio
\beq
\label{brc_resul}
\brc = \left( 1.88^{+0.60}_{-0.38}  \right) \% \, ,
\eeq
where the error corresponds to a variation of $x_c=(m_c/m_b)$ and of
the renormalization scale $\mu$ in the ranges $0.25 \le x_c \le 0.33$
and $0.5 \le \mu/m_b \le 2.0$. The corresponding errors are added in
quadrature. The experimental uncertainty in $\a_s(m_Z)$
has a smaller impact and the errors due to the remaining input parameters
in eq. (\ref{input}) are negligible. The large renormalization scale
dependence of this result is expected to be weakened once the 
current-current
type corrections to the penguin operators are included.

So far, we have considered the {\it charmless hadronic} branching ratio
$\brc$. To obtain the {\it total charmless} branching ratio
$\overline{{\cal B}}(B \to \mbox{no charm})$, one has to add twice
the charmless semileptonic branching ratio
${\cal B}(B \to X_u \ell \overline{\nu}_\ell)$, for $\ell=e$
and $\ell=\mu$ \cite{Nir} (the contribution for $\ell=\tau$,
as well as radiative decay modes can be safely neglected):
\beq
{\cal B}(B \to X_u \ell \overline{\nu}_\ell)=
\left( 0.17 \pm 0.03 \right)\% \times 
\left( \frac{|V_{ub}/V_{cb}|}{0.095} \right)^2 \ .
\eeq
For $|V_{ub}/V_{cb}|=0.095$, we find
\beq
\overline{{\cal B}}(B \to \mbox{no charm})=
\left( 2.22^{+0.60}_{-0.38} \right)\% \ .
\eeq
The experimental result for the total charmless branching ratio
reads
\beq
\overline{{\cal B}}^{\rm{exp}}(B \to \mbox{no charm})=
\left( 0.2 \pm 4.1 \right)\% \ ,
\eeq
obtained in ref. \cite{Neubert98} from CLEO data \cite{CLEO97}.
%%%%
\section{Numerical predictions in the presence of enhanced
    $\boldsymbol{C_8^{\rm{\symbol{101}\symbol{102}\symbol{102}}}}$} 
\label{c8enh}
As discussed in the introduction, the theoretical prediction of the
semileptonic branching ratio and the charm multiplicity are compatible
with the experimental findings if the renormalization scale is allowed
to be as low as $m_b/4$. Both predictions are, however, at the lower
side and therefore an enhancement of the charmless hadronic branching
ratio $\brc$ by new physics would lead to a better agreement. It is therefore
still conceivable that $\brc$ is considerably larger than in the
standard model (SM).

In the SM the initial conditions for $C_{3-6}$ and $C_8$ are generated
at a scale $\mu=O(m_W)$ by the one-loop $bsg$ vertex function.
Due to the fact that the $W$-boson only couples to left-handed quarks,
only chromomagnetic operators proportional to $m_b$ (and $m_s$) are generated.
In extensions of the SM, however, also chromomagnetic operators 
where $m_b$ (or $m_s$) is replaced by the mass of a heavy particle 
propagating in the loop, can be generated \cite{newphys}. Such operators
potentially lead to large contributions to $b \to s g$. 
In the following we will perform a model independent analysis
of the impact of enhanced $C_8$ on $\brc$, emphasizing the role of the
NLL corrections to $b \to s g$. We assume that only chromomagnetic operators
with the same helicity structure as $O_8$ in the SM are generated which can
then be described as a shift in $C_8$. For simplicity, we further assume
that the CKM structure of the new contribution is the same as in the SM,
hence neglecting the possibility of new CP-violating phases, by assuming
the shift in $C_8$ to be real.

\begin{figure}[t]
\begin{center}
\leavevmode
\includegraphics[scale=0.5]{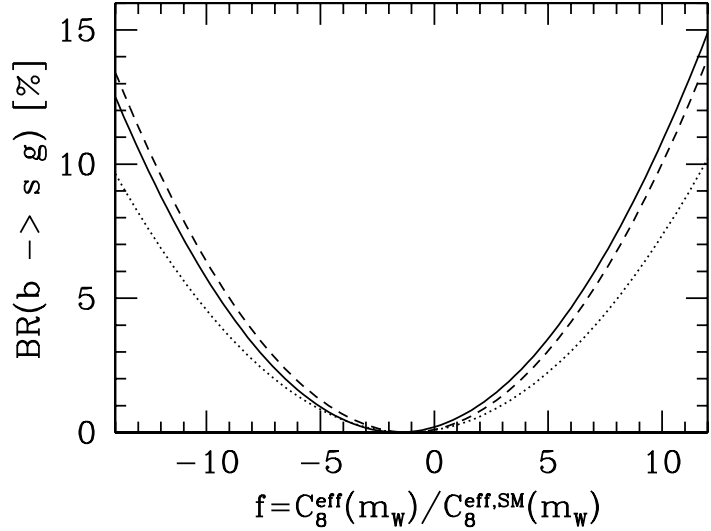}
\end{center}
\caption[f1]{Branching ratio ${\cal B}(b \to s g)$ as a function
of $f=C_8^{\rm{eff}}(m_W)/C_8^{\rm{eff,SM}}(m_W)$. For the exact definition
of $f$, see eq. (\ref{rescal}). The dotted (solid) curve shows the LL (NLL) 
approximation. The dashed curve is obtained by switching off the matrix 
elements of the operators $O_1$ and $O_2$.}
\label{fig:bsgnewmw}
\end{figure}
In fig. \ref{fig:bsgnewmw} we investigate the impact of enhanced
$C_8^{\rm{eff}}(m_W)= C_8^{0,\rm{eff}}(m_W)+ \a_s/(4\p) \, 
 C_8^{1,\rm{eff}}(m_W)$
on the branching ratio for $b \to s g$. In the NLL
approximation for this branching ratio, both, $C_8^{0,\rm{eff}}(m_W)$
and $C_8^{1,\rm{eff}}(m_W)$ enter. In general, it is expected that
the two pieces get different new physics shifts. For purpose of illustration,
we assume however that both pieces are the same 
multiple $f$ of the SM counterparts,
i.e., we assume that
\beq
\label{rescal}
C_8^{0,\rm{eff}}(m_W) = f \,  C_8^{0,\rm{eff,SM}}(m_W) \ ; \quad
C_8^{1,\rm{eff}}(m_W) = f \,  C_8^{1,\rm{eff,SM}}(m_W) \ .
\eeq
The dotted curve shows the LL prediction of ${\cal B}(b \to s g)$
as a function of $f$, while the solid curve shows the NLL prediction.
It is expected that 
for large enhancement factors, the matrix elements of the operators
$O_1$ and $O_2$ become unimportant; this feature is illustrated by the
dashed line, which is obtained by switching off these matrix elements.
The NLL corrections (for large enhancement factors) amount to almost 
$50\%$ of the LL prediction.

\begin{figure}[th]
\begin{center}
\leavevmode
\includegraphics[scale=0.5]{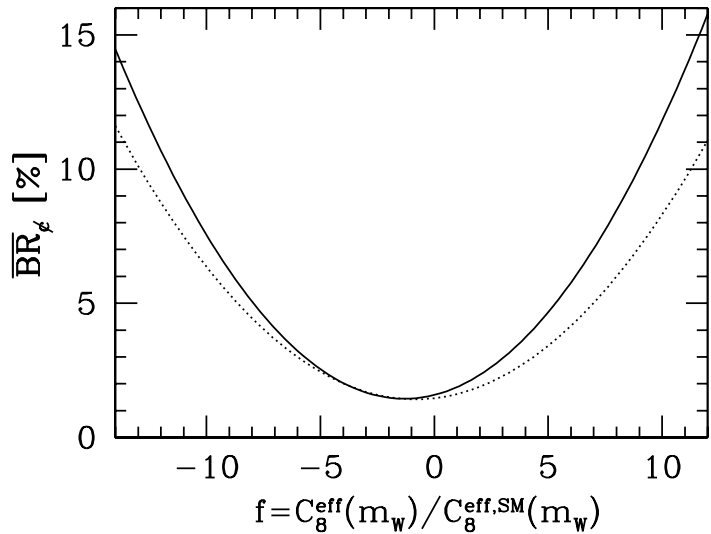}
\end{center}
\caption[f1]{Charmless hadronic branching ratio $\brc$
as a function
of $f=C_8^{\rm{eff}}(m_W)/C_8^{\rm{eff,SM}}(m_W)$. 
For the exact definition
of $f$, see eq. (\ref{rescal}). The dotted (solid) curve includes the LL (NLL) 
approximation for ${\cal B}(b \to g)$. 
The NLL corrections to 
the decay modes with three quark in the final state (see ``approx2''
in section \ref{charmless}) are included in both cases.}
\label{fig:charmlessnew}
\end{figure}
In fig. \ref{fig:charmlessnew},
the impact of enhanced $C_8$ on the charmless hadronic
branching ratio $\brc$ is illustrated. The dotted curve includes 
the NLL corrections to 
the decay modes with three quark in the final state and 
the LL result for ${\cal B}(b \to s g)$ (see ``approx2''
in section \ref{charmless}), while the solid curve also includes
the NLL corrections to   ${\cal B}(b \to s g)$. For a given 
value of $\brc$ (from an ideal measurement), $C_8(m_W)$ can be measured
in principle. To illustrate this, we take the hypothetical
value $\brc=5\%$. The two solutions for the enhancement factor
$f$ are $f=7$ and $f=-9$ when using the dotted curve; including NLL
corrections to $b \to s g$ (solid curve), enhancement factors
with smaller absolute values
do the job, viz.    $f=5$ and $f=-8$.
\section{Summary}
\label{summary}
In this paper we presented a
detailed calculation of 
the $O(\alpha_s)$ virtual corrections to the 
decay width $\G(b \to s g)$.
The most difficult part, the two-loop diagrams associated with the operators
$O_1$ and $O_2$ which from the numerical point of view 
play a crucial role, was obtained by using Mellin-Barnes techniques.
Also complete expressions for the corresponding $O(\alpha_s)$ 
bremsstrahlung corrections to $b \to s g$
were  given.  
The combined result 
is free of infrared and collinear singularities, in accordance with
the KLN theorem.

The renormalized virtually corrected matrix element 
$\bra s g|O_8|b\ket$ contains  logarithms of the form
$\ln(m_f/\mu)$ ($f=u,d,s,c,b$), which for the light flavors
($u,d,s$) represent a special kind of singularity. 
Keeping in mind that these terms originate from the renormalization 
factor $Z_3$ of the gluon field. i.e.,
from gluon self energy diagrams in which these flavors propagate, 
we argued that these singularities cancel against
the logarithms present in the decay rate $\G(b \to s f \bar{f})$ with 
$f=u,d,s$. We therefore included the $O_8$ contribution to
$\G(b \to s f \bar{f})$ for $f=u,d,s$.

Taking into account the existing
next-to-leading logarithmic (NLL) result for the Wilson coefficient
$C_8^{\rm{eff}}$, a complete NLL result for the branching ratio
${\cal B}^{\rm{NLL}}(b \to s g)$ was obtained. Numerically, we found
${\cal B}^{\rm{NLL}}=(5.0 \pm 1.0) \times 10^{-3}$, 
which is more than a factor
of two larger than the leading logarithmic result 
${\cal B}^{\rm{LL}}=(2.2 \pm 0.8) \times 10^{-3}$.
 
We then investigated the impact of these corrections on the 
inclusive charmless hadronic 
branching ratio $\brc$ of $B$-mesons. We found that the NLL corrections
calculated in this paper are of similar importance as NLL
corrections to $b$-quark decay modes with three quarks in the final state,
which were presented by Lenz et al. \cite{Lenz1,Lenz2}. 

Finally, the impact of the NLL corrections to $b \to s g$ on $\brc$
was studied
in scenarios, where the Wilson coefficient $C_8$ is enhanced by new physics.
For a given 
value of $\brc$ (from an ideal measurement), $C_8(m_W)$ can be measured
in principle. To illustrate this, we took the hypothetical
value $\brc=5\%$. The two solutions for the enhancement factor
$f$ are $f=7$ and $f=-9$, using the LL approximation for 
${\cal B}(b \to s g)$; including NLL
corrections to $b \to s g$, somewhat smaller enhancement factors
($f=5$ and $f=-8$) are needed to obtain the hypothetical value
$\brc=5\%$.

\vspace{0.5cm}
\noindent

{\bf Acknowledgments}:
We would like to thank A. Ali, A. Kagan, A. Lenz, P. Minkowski, M. Neubert,
and U. Nierste for helpful discussions.

\newpage

\begin{appendix}
\section{Next-to-leading order Wilson Coefficients}
\label{appendix:a}
In this appendix we present the explicit formulas which 
allow to calculate the Wilson coefficients needed in this paper.

In section \ref{highscale}, we give the results for the Wilson coefficients
at the matching scale $\mu_W$, which is usually taken to be of order
$m_W$. Section \ref{lowscale} is devoted to the
the Wilson coefficients at the scale
$\mu_b$, where $\mu_b$ is of order $m_b$. We give an explicit expression
for $C_8^{\rm{eff}}(\mu_b)$ at NLL, which is new. To make this appendix
self-contained, we also repeat the results for the Wilson coefficients
$C_1(\mu_b)$ and $C_2(\mu_b)$ which are needed 
only to LL precision in our application. 

\subsection{NLL Wilson coefficients at the matching scale $\boldsymbol{\mu_W}$}
\label{highscale}
\noindent 
To give the results for the effective Wilson coefficients $C_i^{\,{\rm eff}}$
at the matching scale $\mu_W$ in a compact form, we
write\footnote{Note that $C_i^{\rm{eff}}(\mu) = C_i(\mu)$ 
by definition for $i=1,...,6$. }:
\begin{eqnarray}
  C^{\,{\rm
      eff}}_i(\mu_W) = C^{0,\,{\rm eff}}_i(\mu_W) + 
  \frac{\alpha_s(\mu_W)}{4\pi}
  C^{1,\,{\rm eff}}_i(\mu_W) \,.
\label{effcoeff}
\end{eqnarray}
The LL Wilson coefficients at this scale are well 
known~\cite{INAMILIM,HW}.
\bea 
 C^{0,\,{\rm eff}}_2(\mu_W)  & = &  1                       \nn \\  
 C^{0,\,{\rm eff}}_i(\mu_W)  & = &  0 
  \hspace*{6.5truecm} (i=1,3,4,5,6)                        \nn \\
  C^{0,\,{\rm eff}}_7(\mu_W)  & = &  
  \frac{x}{24} \, \left[
 \frac{-8x^3+3x^2+12x-7+(18x^2-12x) \ln x}{(x-1)^4} 
                                         \right]    \nn \\
  C^{0,\,{\rm eff}}_8(\mu_W)  & = &   
 \frac{x}{8} \, \left[
 \frac{-x^3+6x^2-3x-2-6x \ln x}{(x-1)^4} \right] \,.
\label{coeffLO}
\eea
The coefficients 
$C_{7}^{0,\rm{eff}}(\mu_W)$ and $C_{8}^{0,\rm{eff}}(\mu_W)$ are
functions of $x=m_t^2/m_W^2$. Note that there is no {\it explicit}
dependence of the matching scale $\mu_W$ in these functions. Whether
there is an {\it implicit} $\mu_W$--dependence via the $t$--quark mass
depends on the precise definition of this mass which has to be
specified when going beyond leading logarithms. If one chooses 
to work with $\overline{m}_t(\mu_W)$, then there is such an
implicit $\mu_W$--dependence of the lowest order Wilson coefficient; in
contrast, when working with the pole mass $m_t$ there is no such
dependence. We choose to express our NLL results in 
terms of the pole mass $m_t$. 

The NLL pieces $C_i^{1,\,{\rm eff}}(\mu_W)$ 
of the Wilson coefficients have an explicit 
dependence on the matching scale $\mu_W$ 
and for $i=7,8$ they also explicitly depend on the actual
definition of the $t$--quark mass. 
Initially, when the heavy particles are integrated out, it is
convenient to work out the matching conditions
$C_i^{1,\,{\rm eff}}(\mu_W)$ for $i=7,8$ 
in terms of ${\overline m_t}(\mu_W)$.
Using eq.~(\ref{polerunning}), it is then   
straightforward to get  the corresponding 
result expressed in terms of the pole mass $m_t$.  
One obtains for $i=1,...,6$:
\bea 
 C_ 1^{1,\,{\rm eff}}(\mu_W) & = &
 15 + 6 \ln\frac{\mu_W^2}{m_W^2} \ ,                            \nn \\
 C_ 4^{1,\,{\rm eff}}(\mu_W) & = &
 E_0 + \frac{2}{3} \ln\frac{\mu_W^2}{m_W^2} \ ,   \nn \\
 C_ i^{1,\,{\rm eff}}(\mu_W) & = &   0 \ , 
  \hspace*{6.0truecm}  (i=2,3,5,6)  
\label{coeffNLO}
\eea
with
\beq
 E_0  = 
 \frac{x (x^2+11x-18)}{12 (x-1)^3}
+\frac{x^2 (4 x^2-16x+15)}{6(x-1)^4} \ln x-\frac{2}{3} \ln x
-\frac{2}{3} \ .                                        
\eeq
For $i=7,8$, we split $C_i^{1,{\rm eff}}(\mu_W)$ into three terms:
\bea
C_{i}^{1,\,{\rm eff}}(\mu_W)                          & = & 
  W_{i} + M_{i} \ln\frac{\mu_W^2}{m_W^2} 
                   + T_{i} 
  \left( \ln\frac{m_t^2}{\mu_W^2} -\frac{4}{3} \right)  \,.
\label{coeffNLOa}
\eea
The first two terms 
$W_{i}$ and $M_{i}$ 
would be the full result when working in terms of
${\overline m}_t(\mu_W)$. $T_{i}$ results when expressing
${\overline m}_t(\mu_W)$ in terms of the pole mass $m_t$ 
in the corresponding lowest
order coefficients. Thus, for $i=7,8$, 
the term $T_{i}$ is obtained as
\beq
\label{Tterm}
T_{i} = 8 \, x \, \frac{\partial C_{i}^{0,\,{\rm eff}}(\mu_W)} {\partial
  x} \,. 
 \eeq 
The explicit form of the
functions $W_{i}$, $M_{i}$ and $T_{i}$ reads
\bea
 W_{7}    & = &
 \frac{-16 x^4 -122 x^3 + 80 x^2 -  8 x}{9 (x-1)^4} \,
  {\rm Li}_2 \left(\!1\!-\!\frac{1}{x} \right)
+\frac{6 x^4+46 x^3-28 x^2}{3 (x-1)^5} \ln^2 x           \nn \\
                   &   &
+\frac{-102x^5-588 x^4-2262 x^3+3244 x^2-1364 x+208}
    {81(x-1)^5} \ln x                                 \nn \\
                   &   &
+\frac{1646x^4+12205x^3-10740x^2+2509x-436}{486(x-1)^4}  \nn \\[1.5ex]
 W_{8}    & = &
 \frac{-4 x^4 +40 x^3 + 41 x^2 + x}{6 (x-1)^4} \,
  {\rm Li}_2 \left(\!1\!-\!\frac{1}{x} \right)
+\frac{ -17 x^3 - 31 x^2}{2 (x-1)^5} \ln^2 x             \nn \\
                   &   &
+\frac{-210 x^5+1086 x^4+4893 x^3+2857 x^2-1994 x+280}
     {216 (x-1)^5} \ln x                              \nn \\
                   &   &
+\frac{737x^4-14102 x^3-28209x^2+610 x-508}{1296(x-1)^4} \nn \\[1.7ex]
 M_{7}   & = &
 \frac{82x^5\!+\!301x^4\!+\!703x^3\!-\!2197x^2\!+\!1319x\!-\!208-
      \! (162x^4\!+\!1242x^3\!-\!756x^2)\ln x}
        {81 (x-1)^5}                                  \nn  \\[1.5ex]
 M_{8}   & = &
\frac{77x^5-475x^4-1111x^3+607x^2+1042x-140+
      \! (918x^3+1674x^2) \,\ln x}
        {108(x-1)^5}                                  \nn \\[1.7ex]
 T_{7}    &=&  \frac{x}{3} \, \left[
\frac{47x^3-63x^2+9x+7-(18x^3+30x^2-24x)
                       \ln x}{(x-1)^5} \right]        \nn \\[1.5ex]
 T_{8}    &=&  2x \, \left[\frac{-x^3-9x^2+9x+1+(6x^2+6x)
                       \ln x}{(x-1)^5} \right] \,.
\label{wcnlosm}  
\eea
The dilogarithm $\mbox{Li}_2(x)$ is defined by
\beq
\mbox{Li}_2(x)=-\int_{0}^{x} \, \frac{dt}{t} \, \ln (1-t) \, .
\eeq
%%%%%%%%%%%%%%%%%%%%%%%%%%%%%%%%%%%%%%%%%%%%
%
\subsection{NLL Wilson coefficients at the low scale $\boldsymbol{\mu_b}$}
\label{lowscale}

\noindent 
The evolution from the matching scale $\mu_W$ down to the
low--energy scale $\mu_b$
is described by the renormalization group equation
\beq
\label{RGE}
 \mu \frac{d}{d\mu} C_i^{{\rm eff}}(\mu) = 
 C_j^{\,{\rm eff}}(\mu) \, \gamma_{ji}^{\,{\rm eff}} (\mu) \,.
\eeq
The initial conditions $C_i^{\,{\rm eff}}(\mu_W)$ for this equation
are given in 
section~\ref{highscale}, while
the anomalous dimension matrix $\gamma_{ij}^{\,{\rm eff}} $ 
up to order $\alpha_s^2$ can be found in ref.{\cite{Misiak97}}. For
completeness we display the result here. The anomalous dimension matrix can
be expanded perturbatively as
\beq 
 \gamma_{ji}^{\,{\rm eff}}(\mu)   = 
   \frac{\alpha_s (\mu)}{4 \pi}  \,
 \gamma_{ji}^{0,\,{\rm eff}}
 + \frac{\alpha_s^2(\mu)}{(4 \pi)^2} \, 
 \gamma_{ji}^{1,\,{\rm eff}} + {\cal O}(\alpha_s^3)\,
\eeq
where matrix  $\gamma_{ji}^{0,\,{\rm eff}}$ is given by
\beq 
\label{gamma0}
 \left\{\gamma_{ji}^{0,\,{\rm eff}}\right\}  = \left(
\begin{array}{rrrrrrrr}
\vspace{0.2cm}
-4       & \f{8}{3}  &       0       &   -\f{2}{9}    &      
 0       &     0     & -\f{208}{243} &  \f{173}{162} \\ 
\vspace{0.2cm}
12       &     0     &       0       &    \f{4}{3}    &     
 0       &     0     &   \f{416}{81} &    \f{70}{27} \\ 
\vspace{0.2cm}
 0       &     0     &       0       &  -\f{52}{3}    &    
 0       &     2     &  -\f{176}{81} &    \f{14}{27} \\ 
\vspace{0.2cm}
 0       &     0     &  -\f{40}{9}   & -\f{100}{9}    & 
\f{4}{9} &  \f{5}{6} & -\f{152}{243} & -\f{587}{162} \\ 
\vspace{0.2cm}
 0       &     0     &       0       & -\f{256}{3}    &     
 0       &    20     & -\f{6272}{81} &  \f{6596}{27} \\ 
\vspace{0.2cm}
 0       &     0     & -\f{256}{9}   &   \f{56}{9}    & 
\f{40}{9}& -\f{2}{3} & \f{4624}{243} &  \f{4772}{81} \\ 
\vspace{0.2cm}
 0       &     0     &       0       &       0        &     
 0       &     0     &     \f{32}{3} &       0       \\ 
\vspace{0.2cm}
 0       &     0     &       0       &       0        &     
 0       &     0     &    -\f{32}{9} &  \f{28}{3}    \\
\end{array} \right) \,, 
\eeq
and in the $\overline{\rm MS}$ scheme with fully anticommuting 
$\gamma_5$, 
$ \gamma_{ji}^{1,\,{\rm eff}}$ is
\beq 
\label{gamma1}
 \left\{\gamma_{ji}^{1,\,{\rm eff}}\right\}  = \left(
\begin{array}{rrrrrrrr}
\vspace{0.2cm}
-\f{355}{9}    & -\f{502}{27}  & -\f{1412}{243}   & -\f{1369}{243}   &    
 \f{134}{243}  &  -\f{35}{162} & -\f{818}{243}    &  \f{3779}{324}   \\ 
\vspace{0.2cm}
 -\f{35}{3}    &   -\f{28}{3}  & -\f{416}{81}     &  \f{1280}{81}    &     
 \f{56}{81}    &   \f{35}{27}  &  \f{508}{81}     &  \f{1841}{108}   \\ 
\vspace{0.2cm}
     0         &        0      & -\f{4468}{81}    & -\f{31469}{81}   &    
 \f{400}{81}   & \f{3373}{108} &  \f{22348}{243}  &  \f{10178}{81}   \\ 
\vspace{0.2cm}
     0         &        0      & -\f{8158}{243}   & -\f{59399}{243}  &   
 \f{269}{486}  & \f{12899}{648}& -\f{17584}{243}  & -\f{172471}{648} \\ 
\vspace{0.2cm}
     0         &        0      & -\f{251680}{81}  & -\f{128648}{81}  &  
 \f{23836}{81} &  \f{6106}{27} & \f{1183696}{729} & \f{2901296}{243} \\ 
\vspace{0.2cm}
     0         &        0      &  \f{58640}{243}  & -\f{26348}{243}  & 
-\f{14324}{243}& -\f{2551}{162}& \f{2480344}{2187}& -\f{3296257}{729}\\ 
\vspace{0.2cm}
     0         &        0      &         0        &         0        &
     0         &        0      & \f{4688}{27}     &         0        \\  
\vspace{0.2cm}
     0         &        0      &         0        &         0        &
     0         &        0      & -\f{2192}{81}    &  \f{4063}{27}    \\
\end{array} \right) \,. 
\eeq

The solution of eq.~(\ref{RGE}), obtained through the
procedure described in~\cite{Buras95}, yields for the 
coefficient $C_8^{\rm{eff}}(\mu_b)$, which we decompose as
\beq
\label{wilsonsplit}
 C_8^{{\rm eff}}(\mu_b) = C_8^{0,\,{\rm eff}}(\mu_b) + 
 \frac{\alpha_s(\mu_b)}{4\pi} \, C_8^{1,\, {\rm eff}}(\mu_b) \,, 
\eeq
the LL term
\bea 
 C^{0,\,{\rm eff}}_8(\mu_b) & = &
  \eta^\frac{14}{23}  C^{0,\,{\rm eff}}_8(\mu_W) 
 + \sum_{i=1}^5 h_i^\prime \,\eta^{a^\prime_i}
  \,C^{0,\,{\rm eff}}_2(\mu_W) \,,
\label{runc80}
\eea
and the NLL contribution
\bea     
 C^{1,\,{\rm eff}}_8(\mu_b) & = & 
  \eta^{\frac{37}{23}} C^{1,\,{\rm eff}}_8(\mu_W) + 
 6.7441 \left( \eta^{\frac{37}{23}} - \eta^{\frac{14}{23}} \right) 
      C^{0,\,{\rm eff}}_8(\mu_W)                         \nn \\[1.05ex] 
                     &   &
 + \sum_{i=1}^8  \left(
   e'_i \,\eta \,C^{1,\,{\rm eff}}_4(\mu_W)
 + (f'_i + k'_i \eta) \,C^{0,\,{\rm eff}}_2(\mu_W)
 +  l'_i \,\eta\, C^{1,\,{\rm eff}}_1(\mu_W) \right) \eta^{a_i} \,.
\label{runc8eff1}
\eea
The symbol $\eta$ is defined as
$\eta=\alpha_s(\mu_W)/\alpha_s(\mu_b)$; 
the vectors $a_i$, $a'_i$ ,$h'_i$, $e'_i$, $f'_i$, $k'_i$ and $l'_i$
read
\bea
\{a_i\} & = & \left\{ \,\tfrac{14}{23}, \,\tfrac{16}{23}, \,\tfrac{6}{23}, \,
-\tfrac{12}{23}, \,0.4086, \,-0.4230, \,-0.8994,
  \,0.1456
\right\}  \quad                                \nn \\
 \{a^\prime_i\} & = &
 \left\{ 
   \, \tfrac{14}{23}, \,0.4086, 
    -0.4230, -0.8994, \,0.1456 
 \right\}                                       \nn \\
 \{h^\prime_i\} & = &
 \left\{ 
   \, \tfrac{313063}{363036}, 
   -0.9135, 0.0873, -0.0571, 0.0209             
 \right\} \nn \\
\{e'_i\} & = & \left\{ \, 2.1399 , \,0 , \,0, \,0, \,-2.6788, \,0.2318 ,
  \,0.3741, \, -0.0670
\right\}  \quad                                \nn \\
\{f'_i\} & = & \left\{ \,-5.8157, \,0, \,1.4062, \,-3.9895, \, 3.2850,
  \,3.6851, \,-0.1424, \, 0.6492
\right\}  \quad                                \nn \\
\{k'_i\} & = & \left\{ \,3.7264, \,0, \,0, \,0, \,-3.2247, \,0.3359, \,0.3812,
  \,-0.2968
\right\}  \quad                                \nn \\
\{l'_i\} & = & \left\{ \,0.2169, \,0, \,0, \,0, \,-0.1793, \,-0.0730,
  \,0.0240, \,0.0113 \right\} \,. \quad
\label{nlorunn} 
\eea

As already mentioned earlier, 
we neglect the contributions of the operators $O_3$,...,$O_6$
in our analysis for $b \to s g$, 
as their Wilson coefficients are rather small.
We therefore only list the results for 
the coefficients
$C_1^{\,{\rm eff}}(\mu_b)$ and $C_2^{\,{\rm eff}}(\mu_b)$, 
which are needed to LL precision only:
 \bea
\label{othercoeff}
 C^{0,\,{\rm eff}}_1(\mu_b) & = & \left( \eta^{\frac{6}{23}} -
\eta^{-\frac{12}{23}} \right) \, 
 C^{0,\,{\rm eff}}_2(\mu_W) \nn \\
 C^{0,\,{\rm eff}}_2(\mu_b) & = & \left( \tfrac{2}{3} \eta^{\frac{6}{23}} +
\tfrac{1}{3} \eta^{-\frac{12}{23}} \right) \, 
 C^{0,\,{\rm eff}}_2(\mu_W) \ .
\eea

When calculating NLL results in the  numerical analysis,
we use the NLL expression for the strong coupling constant:
\beq
\label{aqcd}
\alpha_s(\mu) = \frac{\alpha_s(m_Z)}{v(\mu)} \, 
\left[ 1 - \frac{\beta_1}{\beta_0} \frac{\alpha_s(m_Z)}{4\pi} \,
\frac{\ln v(\mu)}{v(\mu)} \right] \,,
\eeq 
with
\beq
v(\mu) = 1 - \beta_0 \frac{\alpha_s(m_Z)}{2\pi} \, \ln \left(
\frac{m_Z}{\mu} \right) \,,
\eeq
where $\beta_0=\frac{23}{3}$ and $\beta_1=\frac{116}{3}$ (for 5 flavors).
However, for LL results 
we always use the LL expression for $\alpha_s(\mu)$, i.e.,
$\beta_1$ is put to zero in eq.~(\ref{aqcd}).

\section{One-loop functions $\boldsymbol{G_{-1}(\symbol{116})}$ and $\boldsymbol{G_0(\symbol{116})}$}
\label{appendix:b}
In this appendix we give the explicit results for the functions
$G_{-1}(t)$ and $G_0(t)$ needed in eq. (\ref{deltabar}).
Evaluating the integral in eq. (\ref{gi}) for $i=-1,0$, one obtains:
\begin{eqnarray}
  \label{eq:G-1}
  G_{-1}(t) & = & \left\{
    \begin{array}{lr}
      -\frac{\pi^2}{2} + 2 \ln^2\left( \frac{\sqrt t + \sqrt{t-4}}{2} \right) -
      2 i \pi \ln\left( \frac{\sqrt t + \sqrt{t-4}}{2} \right); \qquad & t
      \geq 4 \\[2ex]
      -\frac{\pi^2}{2} - 2 \arctan^2\left( \sqrt{\frac{4-t}{t}} \right) + 2 \pi
      \arctan\left( \sqrt{\frac{4-t}{t}} \right); \qquad &  0 \leq t \leq 4
  \end{array}
  \right.
\end{eqnarray}
\begin{eqnarray}
  \label{eq:G0}
  G_0(t) & = & \left\{ 
  \begin{array}{lr}
    -2 + 2 \sqrt{\frac{t-4}{t}} \ln\left( \frac{\sqrt t + \sqrt{t-4}}{2}
    \right) - i \pi \sqrt{\frac{t-4}{t}};\qquad & t \geq 4 \\[2ex]
    -2 - 2 \sqrt{\frac{4-t}{t}} \arctan\left( \sqrt{\frac{4-t}{t}}  \right) +
    \pi \sqrt{\frac{4-t}{t}};\qquad & 0 \leq t \leq 4
  \end{array}
  \right.
\end{eqnarray}

\end{appendix}

\end{document}